\documentclass[showpacs,aps,twocolumn, nofootinbib]{revtex4}
\usepackage{epsfig}
\usepackage{graphicx}
\usepackage{amsmath,amssymb,amsfonts}
\usepackage{array}
\usepackage{url}
\usepackage{hyperref}
\usepackage{multirow}
\usepackage{float}
\usepackage{lineno}
\usepackage{xspace}
\usepackage{csquotes}
\usepackage{ulem}
\usepackage[usenames,dvipsnames]{color}

\newcommand{\bef}{\begin{figure}}
\newcommand{\eef}{\end{figure}}
\newcommand{\bc}{\begin{center}}
\newcommand{\ec}{\end{center}}

\newcommand{\be}{\begin{equation}}
\newcommand{\ee}{\end{equation}}
\newcommand{\bea}{\begin{eqnarray}}
\newcommand{\eea}{\end{eqnarray}}
\newcommand{\omegav}{\boldsymbol{\omega}}

\def\ba{\begin{eqnarray}}
\def\ea{\end{eqnarray}}
\begin{document}
\title{Rotational susceptibility of a hot and dense hadronic matter}
\author{Bhagyarathi Sahoo$^{1}$}
\author{Kshitish Kumar Pradhan$^{1}$}
\author{Dushmanta Sahu$^{2}$}
\author{Raghunath Sahoo$^{1}$\footnote{Corresponding author email: Raghunath.Sahoo@cern.ch}}
\affiliation{Department of Physics, Indian Institute of Technology Indore, Simrol, Indore 453552, India}
\affiliation{$^{2}$ Instituto de Ciencias Nucleares, Universidad Nacional Autónoma de México, Apartado Postal 70-543, México Distrito Federal 04510, México}

\begin{abstract}
We study the effect of global rotation on rotational susceptibilities ($\chi^{(1)}_{\rm \omega}$, $\chi^{2}_{\rm \omega}$, etc.), which quantify how much the system responds to small angular velocities, in a hadron resonance gas produced by ultra-relativistic heavy ion collisions. The higher-order rotational susceptibilities and their ratios are estimated in the presence and absence of baryon chemical potential ($\mu_{\rm B}$) in the system. The effect of particle spin ($s$) and system size ($R$) on the first- and second-order rotational susceptibility is explored. To consider a more realistic scenario, the effect of interactions between hadrons is taken into account by considering van der Waals-like interactions, which include both attractive and repulsive interactions. To validate our results, a comparison with the ideal HRG as a baseline and a 3-flavour NJL model is shown. A nuclear liquid-gas phase transition, which is the characteristic feature of the van der Waals hadron resonance gas model, absent in an ideal hadron gas model, is probed via global rotation.

\end{abstract}
\date{\today}
\maketitle
\section{Introduction}
\label{intro}
One of the interesting areas of research in high-energy physics is understanding the quantum chromodynamics (QCD) phase diagram. Both temperature ($T$) and baryon density ($\mu_{\rm B}$)  play crucial roles in characterizing the phase transition from hadrons to a strongly interacting thermalized partonic matter, known as quark-gluon plasma (QGP). At vanishing or small baryochemical potential and very high temperature, the phase transition is of crossover type~\cite{Aoki:2006we, Ding:2015ona, Ratti:2018ksb, Bazavov:2019lgz}. However, at very high baryochemical potential and smaller temperatures, we expect a first-order phase transition to happen~\cite{Fukushima:2013rx, Fukushima:2010bq, Fischer:2018sdj, Fu:2022gou}. Both kinds of these phase transition lines are expected to meet at a point called the critical endpoint (CEP)~\cite{Stephanov:1998dy, Stephanov:2004wx, Asakawa:2015ybt}. Although the existence of CEP is yet to be confirmed by experiments, it has been a very hot topic of interest in the community. At the CEP, the static thermodynamic quantities show significant fluctuations due to large correlation lengths~\cite{Stephanov:1998dy, Stephanov:1999zu}. The conserved charges associated with the susceptibilities have very different values in the confined and deconfined medium, thus allowing us to distinguish between the two phases. The measurements of the moments of conserved charges, such as baryon number ($B$), strangeness ($S$), and charge ($Q$), which are linked to their corresponding susceptibilities, can help us understand the critical point and subsequently the full QCD phase diagram~\cite{Bazavov:2020bjn, Li:2023mpv}. Thus, a theoretical study of such conserved quantities to understand the systems formed in ultra-relativistic collisions becomes necessary. Many studies have been conducted in this area by taking the hadron resonance gas (HRG) model~\cite{Garg:2013ata} and its modifications such as the excluded volume hadron resonance gas (EVHRG) model~\cite{Bhattacharyya:2013oya, Fu:2012zzc}, mean-field hadron resonance gas (MFHRG) model~\cite{Pal:2023zzp}, and van der Waals hadron resonance gas (VDWHRG) approach in the hadronic phase~\cite{Samanta:2017yhh, Vovchenko:2016rkn}. Although the HRG model agrees well with the lattice QCD (lQCD) results at a low-temperature regime for estimating thermodynamic quantities, it fails to match with lQCD while estimating the higher-order susceptibilities~\cite{Vovchenko:2016rkn}. An excluded volume HRG or the mean-field HRG model does a better job compared to an ideal HRG model. However, with both attractive and repulsive interactions, the van der Waals HRG model can give us better estimations of these susceptibilities while matching well with the lQCD data up to $T\sim180$ MeV~\cite{Samanta:2017yhh, Vovchenko:2016rkn}.

Recently, a new window has opened up in the field of high-energy collisions related to the initial orbital angular momentum created during the collisions and the subsequent vorticity formation and evolution. It is expected that a huge amount of vorticity is produced in peripheral heavy-ion collisions, along with a substantial magnetic field being produced due to the charged spectators. In a thermalized medium, this vorticity causes spin polarization of the particles in the system. Thus, the initial angular momentum of the system can be connected to the observation of spin polarization of various vector mesons and hyperons~\cite{STAR:2017ckg, Sahoo:2024yud, Sahoo:2024egx, Sahoo:2023oid, Sahoo:2025bkx}. This vorticity or rotation in the medium brings new insights into the area of heavy-ion collisions and, more importantly, to the study of phase transitions. Like temperature and baryon density, rotation ($\omega$) can also help in the deconfinement of partons \cite{Mameda:2023sst}. Thus, the 2D QCD phase diagram changes into a 3D one, with rotation in the 3rd axis. Exploring this new avenue for a better understanding of the medium formed in heavy-ion collisions is fascinating and timely.\\

Apart from this, there are various other ways in which vorticity can be generated in the system. For example, the jet-like fluctuations in the expanding fireball can induce a smoke-loop-type vortex around fast-moving particles~\cite{Betz:2007kg}. In addition, the inhomogeneous expansion of the fireball can also generate vorticity~\cite{Xia:2018tes, Jiang:2016woz, Wei:2018zfb, Becattini:2017gcx}. Moreover, vorticity can also be generated due to the Einstein-de Haas effect, which states that the magnetization in the QCD matter will induce a finite rotation in the system~\cite{Einstein:1915}. The finite viscosity present in the system can also generate vorticity and vice-versa~\cite{Sahoo:2023xnu}. Thus, even if the system is not produced from a non-central collision, there are still various ways in which rotation can be introduced into the medium. The strength of the vorticity field produced in the heavy-ion collisions can be estimated from the average global hyperon polarization measurement of $\Lambda$ and $\bar{\Lambda}$ hyperons using a non-relativistic statistical thermal model~\cite{Becattini:2016gvu}. Recently, the slope of the directed flow of hadrons has been used to get an estimate of the strength of the vorticity field~\cite{Jiang:2023vxp}. Moreover, this vorticity field induces the flow of axial current along the axis of rotation known as the chiral vortical effect (CVE), and collective excitations are called chiral vortical wave (CVW)~\cite{Gao:2012ix}. Thus, it becomes essential to understand and study the effect of the rotation of the medium formed in high-energy collisions.

In the presence of rotation, the general Euler's thermodynamic relation changes, with the rotation chemical potential playing an important role~\cite{Becattini:2010, Florkowski:2017ruc}. According to Noether's theorem, the conserved quantum number associated with the rotational symmetry of the system is the total angular momentum, which is globally conserved. Near the QCD phase transition, the medium undergoes a significant change in its degrees of freedom, and consequently in its response to rotation, reflecting a redistribution of angular momentum between the partonic and hadronic phases. Although total angular momentum is conserved globally, the medium's capacity to store and transport angular momentum via spin and vorticity changes. This potentially offers a novel probe of the system's response to rotation and can be quantified by the rotational susceptibility. It is an analogue of magnetic susceptibility, which is defined as the derivative of the free energy of the medium with respect to the rotation.
The rotational (or vortical) susceptibility estimates the spin polarization of quarks and antiquarks in a finite-density QCD matter~\cite{Li:2020eon}. The first- and second-order quark spin polarization (first and second-order rotational susceptibility), correlation, and fluctuations have been estimated using the three-flavor rotating Nambu–Jona–Lasinio (NJL) model~\cite{Sun:2021hxo, Chen:2024hki}. The vortical susceptibility of finite-density QCD matter in the chirally broken phase was calculated in Ref.~\cite{Aristova:2016wxe}. Moreover, the study of rotational susceptibility has implications in low-energy solid-state physics~\cite{Gadagkar:2012, Fur:2012}. Although the rotational susceptibility calculation on the lattice simulation has not been explored yet, the magnetic susceptibility is estimated in Ref.~\cite{Bali:2012jv}.

To study this in the hadronic phase, one can use the hadron resonance gas (HRG) model, where hadrons are treated as point-like particles without any interactions among them.  This simplistic model can explain the lQCD data of the thermodynamic behavior of the matter produced in high-energy collisions at low temperatures. However, after $T \simeq 150$ MeV, the hadrons start to melt, and the HRG model estimations deviate from the lQCD data~\cite{Karsch:2003vd}. The HRG model is widely used in the literature, albeit with its various assumptions. However, a more realistic case is incorporated by including both attractive and repulsive interactions among the hadrons in the so-called van der Waals hadron resonance gas (VDWHRG) model~\cite{Vovchenko:2016rkn}. In our previous work~\cite{Pradhan:2023rvf}, we have studied the effect of global rotation on the thermodynamic properties of an interacting hadron resonance gas, where we showed how rotation affects the liquid-gas phase transition. In this work, we estimate the first, second, and higher-order rotational susceptibilities in both the ideal and van der Waals hadron resonance gas models. We examine the effect of both attractive and repulsive interactions in a hadron gas under rotation. The higher-order rotational susceptibilities and their ratios are estimated at zero and finite baryon chemical potential. We find that these ratios show peculiar characteristics near the liquid-gas phase transition temperature and baryon chemical potential, which is encoded in the VDWHRG model and probed via global rotation. Further, these ratios are found to be sensitive to rotation and follow a particular ordering among themselves.

The structure of this article is as follows. In section \ref{formulation}, we establish the methodology that we use for the calculation of rotational susceptibilities within the scope of a rotating VDWHRG model. In the next section \ref{res}, we discuss the findings, and finally, in section \ref{sum}, we summarise our results and highlight the possibility of future study along these directions.

\section{Formulation}
\label{formulation}

\subsection{Rotating HRG model with van der Waals interaction}
\label{ss1}
The single-particle phase-space distribution function $f({\bf x},{\bf p}, \omegav)$ for an ideal relativistic system of massive fermions (bosons) with half-integral (integral) spin ($s$) rotating with angular velocity $\omegav$ is obtained in our previous study~\cite{Pradhan:2023rvf}.  The phase-space distribution function $f({\bf x},{\bf p}, \omegav)$ is given by  

\begin{equation}\begin{split}
f({\bf x},{\bf p}, \omegav) &=  \frac{\exp{[({\bf p} \cdot {\bf v})/ T ]}}{\exp{[(E_{i} - \mu_{i})/T]} \pm 1}  \frac{\sinh(s+\frac{1}{2})\frac{\omega}{T}}{\sinh(\frac{\omega}{2T})},
\end{split}
\label{fxpw}
\end{equation}  

where $E_i = \sqrt{p^2 + m_i^2}$ is the  energy and $m_{i}$ being the mass of the $i^{th}$ hadron. Here, $\bf v$ is the velocity of the particle. The $\pm$ sign corresponds to fermions and bosons, respectively. Here, $\mu_{i}$ is the total chemical potential of the system, which can be further expanded into 
\begin{equation}
\mu_i = B_i\mu_B + S_i\mu_S +Q_i\mu_Q
\label{mub}
\end{equation}
where $\mu_{\rm B}$, $\mu_{\rm S}$, and $\mu_{\rm Q}$ are the baryon chemical potential, strangeness chemical potential, and electric charge chemical potential, respectively. With $B_i$, $S_i$, and $Q_i$ denoting the baryon number, strangeness, and electric charge quantum number of the $i^{th}$ hadron, respectively. \\

To derive Eq.~(\ref{fxpw}), we assume a system of relativistic gas, rotating with a constant angular velocity vector $\omegav$.  To have a constant angular velocity vector, $\omegav$, we can define the rigid velocity, $\bold{v} = \omegav \times \bold{x}$ for a system of size $x$.
A similar phase space distribution function is obtained for classical rotating gas in Ref.~\cite{Becattini:2009wh}. The non-relativistic thermodynamic equilibrium in a rotating system can be defined when the rotation is assumed to be rigid \cite{Becattini:2007nd, Becattini:2009wh, landau}. In the relativistic system, another constraint of $|\omegav \times \bold{x}| \ll 1$ (in natural units) is added to respect the causality condition. For a proper macroscopic system (and in fact, for the majority of practical reasons), the ratio of $\omega$ to $T$ is very small, i.e., $\frac{\hbar \omega}{k_{B}T} \ll 1$. The range of rotation $\omega$ and temperature $T$ values considered in this study satisfies this condition.  Under this condition, Eq.~\ref{fxpw} is obtained by expanding to leading order in $\frac{\omega}{T}$ (in natural units $\hbar = k_{B}$  = 1). Therefore, the higher-order contribution of $\frac{\omega}{T}$ to the phase space distribution function and hence to the resulting observable is expected to be negligible. In this study, we assume a rotation to be a rigid rotation; for this, we take the system size, $x=R$, to be 5 fm throughout the work, which is a safe assumption for the hadronic system produced in high-energy collisions. But it is important to note that the value of the chosen $R$ will have a significant impact on the results. The values of $\omega$ are chosen such that they respect the causality condition ($\omega R < 1$).  Also, for the considered values of $\omega$, the Lorentz factor is found to be close to unity and will have a negligible impact on the results. Hence, it is neglected in the current formalism. \\

The ideal thermodynamic pressure variable for the $i^{th}$ particle ($P_i^{id}$) for a rotating system can be obtained from the phase-space distribution function $f({\bf x},{\bf p}, \omegav)$ in Eq. (~\ref{fxpw}) and is given by;

\begin{equation}
P_i^{id} = \frac{g_i}{2\pi^2} \int_{0}^{\infty} p^2 dp\ \frac{p^2}{3E_i} f({\bf x},{\bf p}, \omegav).
\label{idP}
\end{equation}

Here, $g_i$ is the degeneracy of $i^{th}$ hadron. By summing the pressure over all hadron species, the total pressure of the hadron resonance gas model can be obtained as follows

\begin{equation}
P^{id} = \sum_i P^{id}_i.
\label{totP}
\end{equation}

It is worth mentioning that the ideal HRG model does not incorporate any interactions among the hadrons. Therefore, the ideal HRG model is unable to explain different thermodynamic quantities, and conserved charge susceptibilities results estimated from the first principle based on lQCD calculations at high temperatures and baryon densities. With time, several statistical models have been developed to include the interactions in the medium. Now, to start with, we use the van der Waals equation of state (EoS) in the canonical ensemble representation, which can be written as \cite{Vovchenko:2016rkn}
\begin{equation}
\label{vdweq}
\Bigg( P + \bigg(\frac{N}{V}\bigg)^{2}a\Bigg)\big(V-Nb\big) =  N T
\end{equation}
where $a$ and $b$ are the VDW parameters, and describe the attractive and repulsive interactions among the hadrons, respectively. The $P$, $V$, $T$, and $N$ represent the pressure, volume, temperature, and the number of particles in the system, respectively.

Writing the above Eq. (\ref{vdweq}) in terms of the number density, $n \equiv N/V,$ we have
\begin{equation}
\label{vdweq2}
P(T,n) = \frac{nT}{1-bn}- an^{2},
\end{equation}

The two correction terms in the above Eq. (~\ref{vdweq2}) compared to the ideal case ($P(T,n) = nT$) are due to the repulsive and attractive interactions. The repulsive interaction (first term) is incorporated via the excluded volume correction method by changing the total volume V to an effective volume that is accessible to particles. It is taken care of by using the appropriate volume parameter $b = 16\pi r^3/3$, where $r$ is the particle's hardcore radius. We use $r_{M}=0.2$ fm for mesons, and $r_{B,(\bar{B})}=0.62$ fm for (anti)baryons~\cite{Pradhan:2023rvf, Sahoo:2023vkw, Albright:2014gva, Sarkar:2018mbk}. In contrast, the attractive interaction (second term) is set to $a=0.926$ GeV fm$^{3}$ \cite{Sarkar:2018mbk, Pradhan:2023rvf}. The considered hardcore radius of baryons is based on a chi-square fit to the lattice calculation of both pressure and the trace anomaly (or interaction measure). In Ref.~\cite{Albright:2014gva}, the authors provide insights about the hard core radius of protons and neutrons, which is found to be 0.62 $\pm$ 0.04 fm. There are several methods for extracting the VDW parameters. These can be estimated by reproducing the ground state of the nuclear matter \cite{Vovchenko:2015vxa}, or by fitting lQCD results for different thermodynamic quantities like energy density and pressure, etc. \cite {Samanta:2017yhh, Sarkar:2018mbk}. In this study, the VDW parameters used are obtained by contrasting the lattice equation of state for QCD matter at finite chemical potentials~\cite{Sarkar:2018mbk}. The Eq. (\ref{vdweq2}) reduces to the EVHRG equation of state for $a = 0$, where only repulsive interactions are included. And it reduces to the ideal HRG case for both $a = 0$ and $b = 0$.

For the Grand Canonical Ensemble (GCE), a similar procedure can be followed, and the VDW equation of state can be expressed as follows \cite{Vovchenko:2015vxa, Vovchenko:2015pya} 
\begin{equation}
\label{vdwp}
P(T,\mu) = P^{id}(T,\mu^{*}) - an^{2}(T,\mu),
\end{equation}
where, $P(T,\mu)$, and $P^{id}(T,\mu)$ is the VDW, and ideal pressure, respectively. When there is no interaction in the system, the VDW pressure reduces to the ideal HRG pressure. The particle number density of the VDWHRG model, $n(T,\mu$) is given by
\begin{equation}
\label{vdwn}
n(T,\mu) = \frac{\sum_{i}n_{i}^{id}(T,\mu^{*})}{1+b\sum_{i}n_{i}^{id}(T,\mu^{*})}.
\end{equation}
Here, $i$ runs over all hadrons and resonances in the interacting system, and $\mu^{*}$ is the modified chemical potential due to the interaction present in the system. This $\mu^{*}$ can be parametrized as follows 
\begin{equation}
\label{mustar1}
\mu^{*} = \mu - bP(T,\mu) - abn^{2}(T,\mu) + 2an(T,\mu).
\end{equation}
Using Eqs. (\ref{vdwp}) and (\ref{vdweq2}), $\mu^{*}$ can also be written as
\begin{equation}
\label{mustar2}
\mu^{*} = \mu - \frac{bn(T,\mu)T}{1-bn(T,\mu)} + 2an(T,\mu).
\end{equation}

The VDWHRG model, as originally proposed, incorporates interactions limited to all baryon or anti-baryon pairs~\cite{Samanta:2017yhh, Vovchenko:2015vxa, Vovchenko:2015pya, Vovchenko:2016rkn}. The interaction between baryon-antibaryon couples was neglected by assuming the short-range interactions are dominated by annihilation processes~\cite{Andronic:2012ut, Vovchenko:2016rkn}. All meson-related interactions, including meson-meson and meson-(anti)baryon interactions, were previously disregarded because the inclusion of these interactions suppresses the thermodynamic quantities in comparison with lQCD data at the crossover region~\cite{Vovchenko:2016rkn}. Nevertheless, a more realistic formalism that considers meson-meson interactions was developed by selecting the VDW parameters that best suit the lQCD data by assuming a hard-core radius $r_M$ for mesons~\cite{Sarkar:2018mbk}. Consequently, the total pressure of the VDWHRG model is denoted as; 

\begin{equation}
\label{finalp}
P(T,\mu) = P_{M}(T,\mu) + P_{B}(T,\mu) + P_{\bar{B}}(T,\mu),
\end{equation}
where, $P_{M}(T,\mu),  P_{B(\bar B)}(T,\mu)$ denotes the pressure contributions due to mesons $M$ and (anti)baryons $B (\bar{B})$, respectively. The pressure in the VDWHRG model is related to the ideal pressure via the relation
\begin{equation}
\label{mesonp}
P_{M}(T,\mu) = \sum_{i\in M}P_{i}^{id}(T,\mu^{*M}),       
\end{equation}
\begin{equation}
\label{baryonp}
P_{B}(T,\mu) = \sum_{i\in B}P_{i}^{id}(T,\mu^{*B})-an^{2}_{B}(T,\mu),
\end{equation}
\begin{equation}
\label{antip}
P_{\bar{B}}(T,\mu) = \sum_{i\in \bar{B}}P_{i}^{id}(T,\mu^{*\bar{B}})-an^{2}_{\bar{B}}(T,\mu).
\end{equation}
Because of interaction, the $\mu^{*M}$, $\mu^{*B}$ and $\mu^{*\bar B}$ represent the modified chemical potentials for mesons, baryons and anti-baryons, respectively, and can be determined from Eq. (~\ref{mub}) and Eq. (~\ref{mustar1}) as; \begin{equation}
\label{mumeson}
\mu^{*M} = -bP_{M}(T,\mu),
\end{equation}
\begin{equation}
\label{mubaryon}
\mu^{*B(\bar B)} = \mu_{B(\bar B)}-bP_{B(\bar B)}(T,\mu)-abn^{2}_{B(\bar B)}+2an_{B(\bar B)},
\end{equation}
where $n_{M}$, $n_{B}$, and $n_{\bar B}$ denote the modified number densities for mesons, baryons, and anti-baryons, respectively, which are given by
\begin{equation}
\label{nmeson}
n_{M}(T,\mu) = \frac{\sum_{i\in M}n_{i}^{id}(T,\mu^{*M})}{1+b\sum_{i\in M}n_{i}^{id}(T,\mu^{*M})},
\end{equation}
\begin{equation}
\label{nbaryon}
n_{B(\bar B)}(T,\mu) = \frac{\sum_{i\in B(\bar B)}n_{i}^{id}(T,\mu^{*B(\bar B)})}{1+b\sum_{i\in B(\bar B)}n_{i}^{id}(T,\mu^{*B(\bar B)})}.
\end{equation}

It is important to note that the van der Waals parameters $a$ and $b$ used in this study are obtained in the absence of rotation in the medium. This can be considered a reasonable assumption in the first-order approximation, particularly in the small angular velocity regime where linear response arguments are expected to apply. In principle, rotation can modify local density distributions, induce centrifugal gradients, and alter the effective interaction environment of the medium. Such effects may lead to a rotation dependence of thermodynamic quantities and, consequently, of effective interaction parameters. This consideration becomes especially relevant for higher-order rotational susceptibilities, which can amplify underlying modeling assumptions. Thus, in principle, these parameters should vary as a function of rotation. However, as it is non-trivial to have $a$ and $b$ as functions of $\omega$, we neglect the dependency in the current study.


\subsection{Fluctuation and conserved charge}
\label{flucor}
The nth-order susceptibility is defined as 
\begin{equation}
\chi_{i}^{(n)}= \frac{\partial^n(P(T,\mu_i)/T^4 )}{\partial (\frac{\mu_{i}}{T})^n}.
\label{eq2}
\end{equation}
Here, $\mu_i$ is the chemical potential for the corresponding conserved charges. We take $i$ to be baryon number ($B$),
electric charge ($Q$) and strangeness ($S$). The $n=1,2,3,4$ susceptibilities are related to the mean, variance, skewness, and kurtosis of the distribution of the conserved charges. 

In the presence of rotation, the standard Euler's thermodynamic relation is modified into,
\begin{equation}
\varepsilon + P = Ts + \mu n + \omega \rm w
\label{eq1}
\end{equation}

where, $\omega$ is the rotation and $\rm w$ is the  angular momentum density. Equation (\ref{eq1}) indicates that, in a rotating system, the angular velocity $\omega$ acts as the thermodynamic variable conjugate to the conserved angular momentum. The correlations and fluctuations of any conserved charges are intrinsically linked to the corresponding susceptibility. Thus, we define a susceptibility associated with the rotation of the system, which can be written as,

\begin{equation}
\chi_{\rm \omega}^{(n)}= \frac{\partial^n[P(T, \mu_i, \omega)/T^4 ]}{\partial (\frac{\omega}{T})^n}.
\label{eq3}
\end{equation}

The first-order derivative of the pressure with rotation is called the angular momentum and is defined as follows;
\begin{equation}
J = \frac{\partial P(T, \mu_i, \omega)}{\partial \omega}.
\label{eq4}
\end{equation}

For a rigid body rotation of the system, total angular momentum is conserved. On the other hand, the second-order derivative of the pressure with respect to the angular velocity, equivalently, the first-order derivative of the angular momentum with respect to the angular velocity, defines the moment of inertia, \(I\), as

\begin{equation}
I = \frac{\partial J}{\partial \omega} = \frac{\partial^2 P(T, \mu_i, \omega)}{\partial \omega^2}.
\label{eq4}
\end{equation}
The moment of inertia $I$ is equivalent to the second-order rotational susceptibility  $\chi^{(2)}_\omega$.


Unlike conserved charge susceptibilities (e.g., for baryon number, electric charge, or strangeness), $\chi_\omega$ does not originate from a fundamental symmetry. Instead, it is analogous to the magnetic susceptibility $\chi_M$, which is also defined as a statistical response function tied to fluctuations of a non-conserved quantity, known as the magnetization. Both $\chi_M$ and $\chi_\omega$ reflect the system's intrinsic tendency to respond to weak external fields (magnetic or rotational) through thermodynamic fluctuations. Thus, studying the susceptibility under the influence of rotation can provide insights into the correlations and fluctuations of angular momentum.

By using the formalism mentioned above, we estimate the rotational susceptibilities and their ratios for a rotating hadron gas with van der Waals interaction.

\section{Results and Discussion}
\label{res}
\begin{figure*}
\includegraphics[scale=0.37]{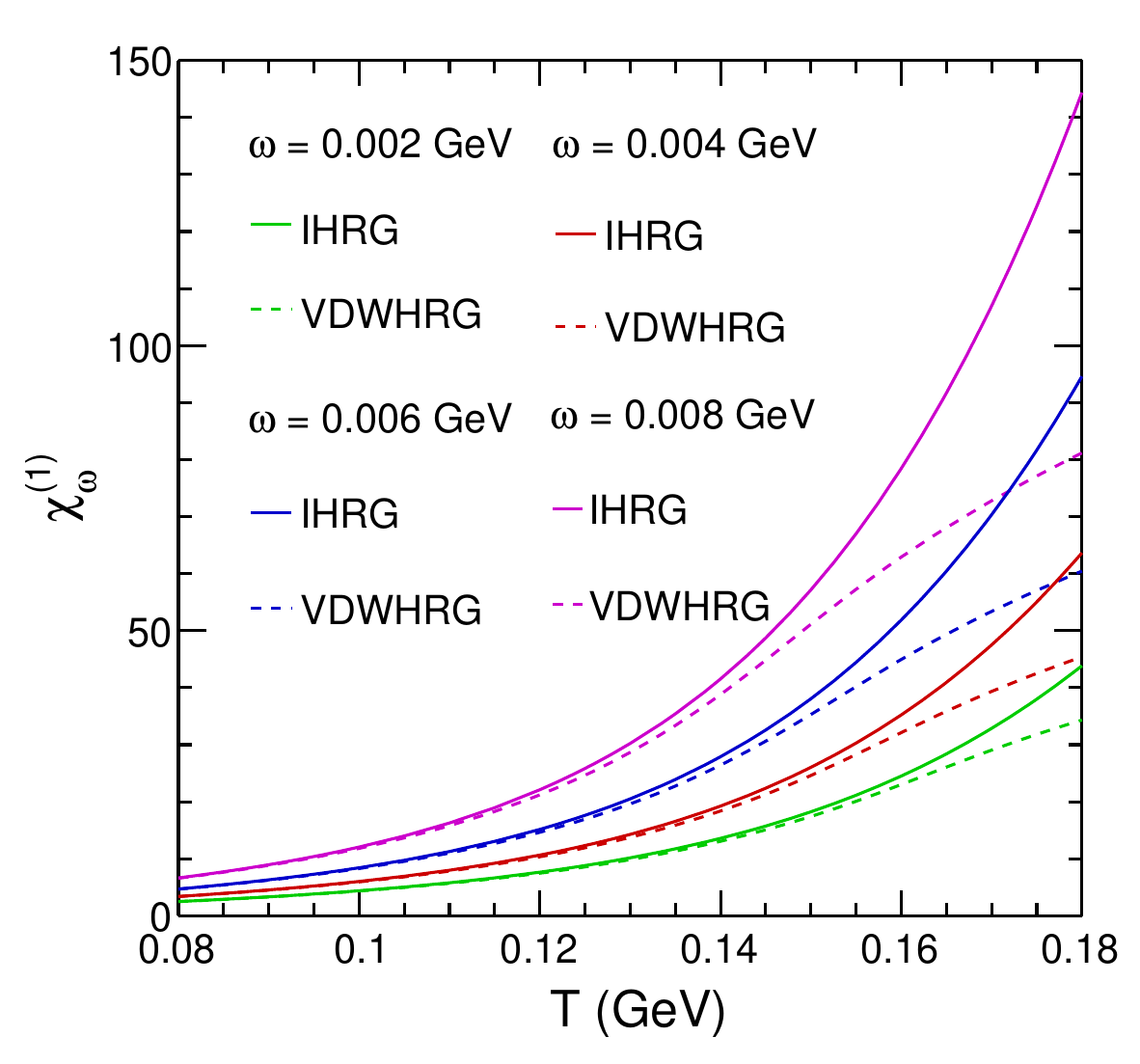}
\includegraphics[scale=0.37]{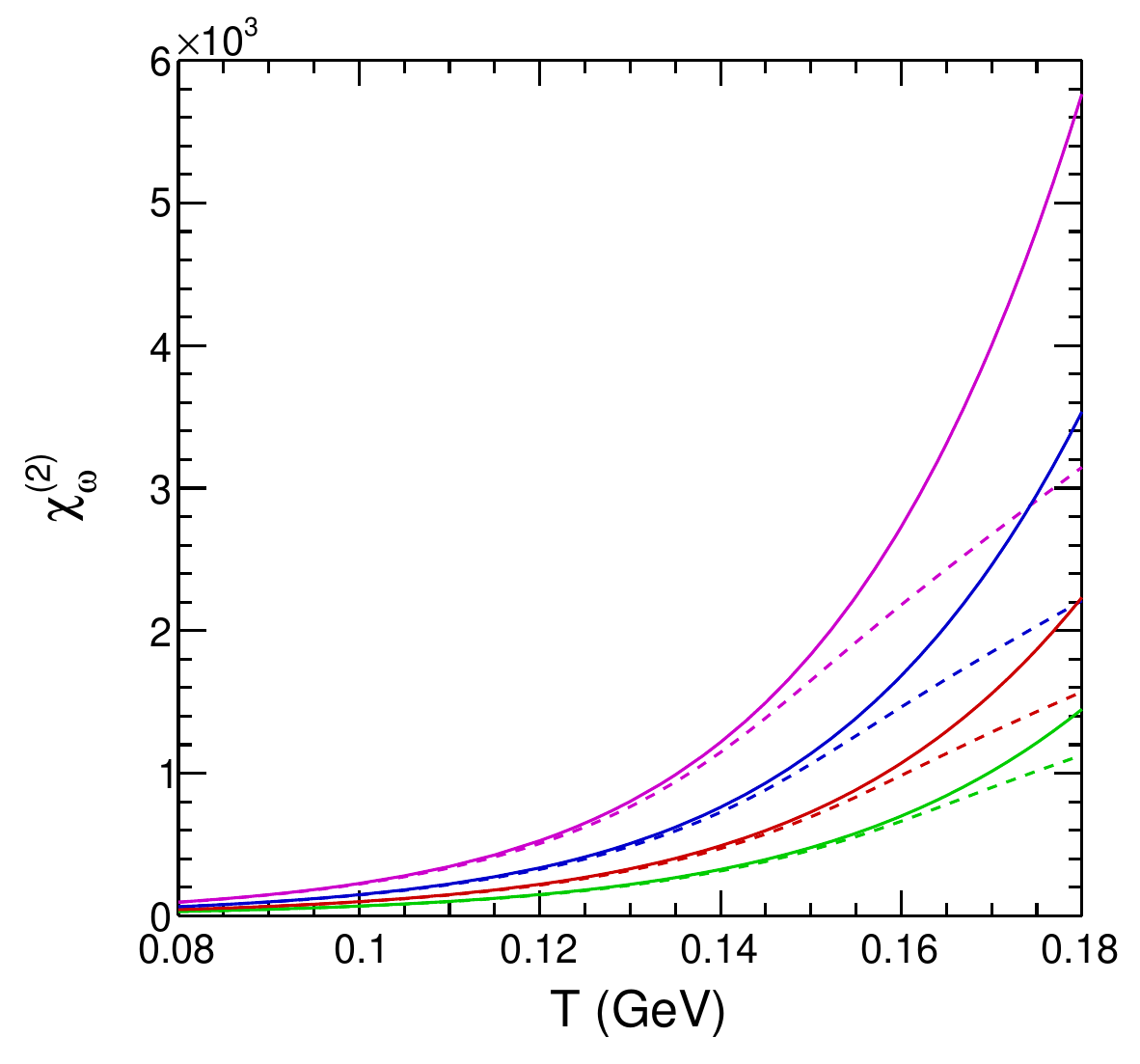}
\includegraphics[scale=0.37]{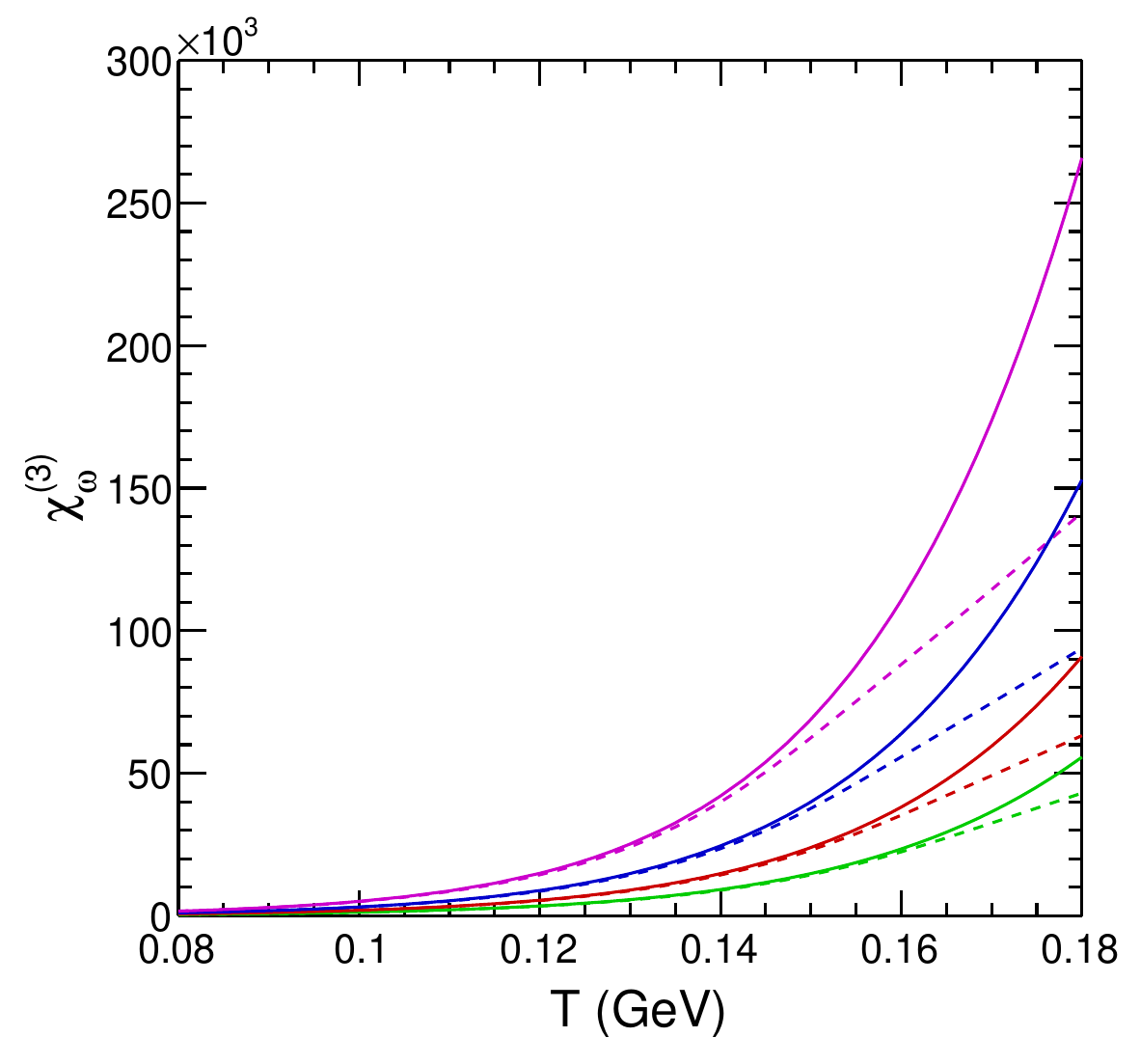}
\includegraphics[scale=0.37]{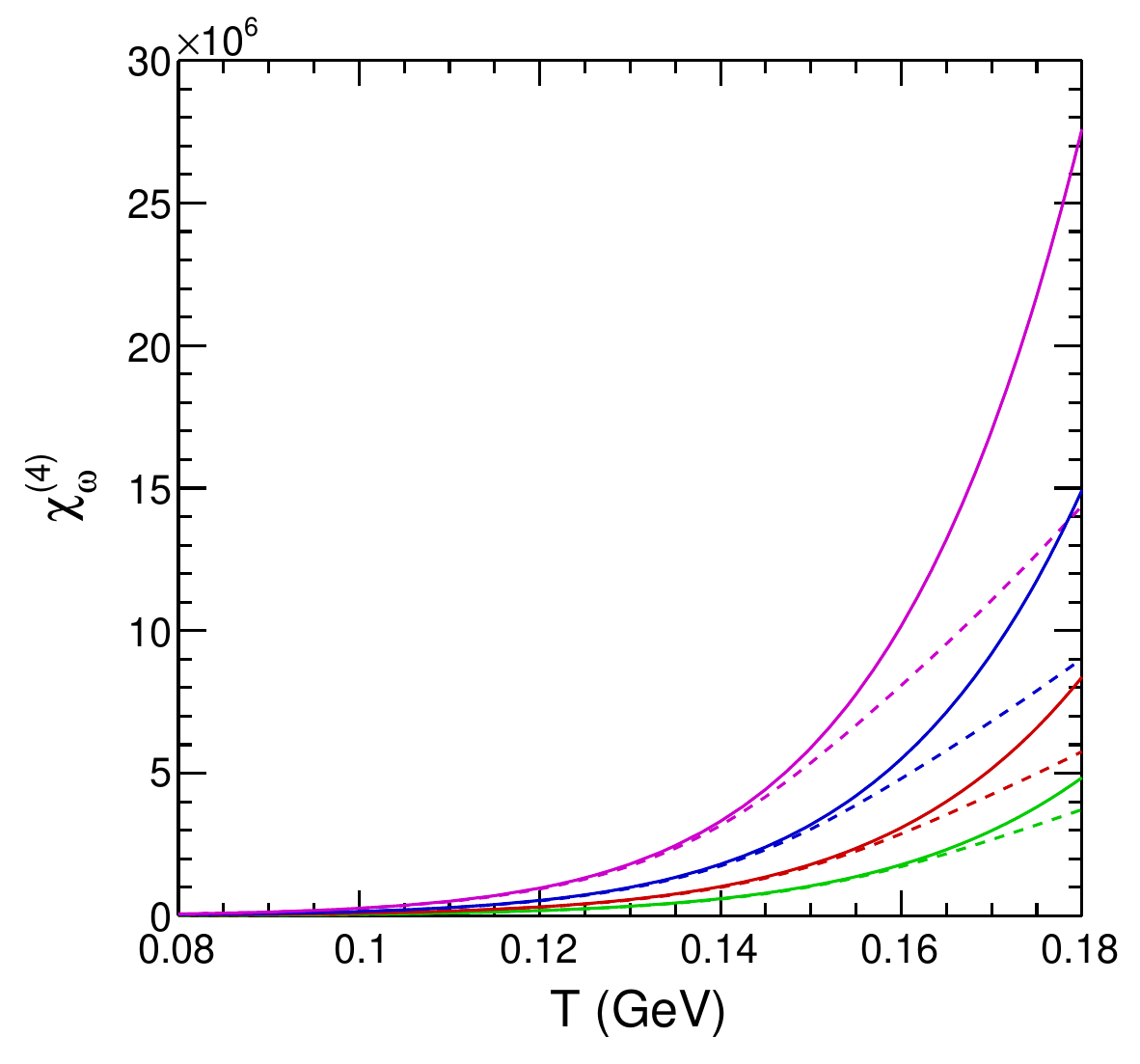}  
\caption{(Color online) The rotational susceptibilities in the ideal HRG (VDWHRG) model are shown in a solid line (dotted line). The variation (from left to right and downwards) of $\chi^{(1)}_\omega$, $\chi^{(2)}_\omega$, $\chi^{(3)}_\omega$, $\chi^{(4)}_\omega$ as a functions of temperature at zero baryochemical potential ($\mu_{\rm B}$ = 0 GeV), for $\omega$ = 0.002 GeV (green), $\omega$ = 0.004 GeV (red), $\omega$ = 0.006 GeV (blue), and $\omega$ = 0.008 GeV (magenta). The observed hierarchy in the rotational susceptibilities reflects the increasing sensitivity of higher-order susceptibilities  to temperature and rotation in both the ideal HRG (VDWHRG) model.}
\label{fig:susvsC}
\end{figure*}

In this section, we will examine the findings of this study. It is noteworthy to mention that all the results obtained in this study at $\mu_{\rm Q} =$ 0 and $\mu_{\rm S} =$ 0. This is a reasonable approximation at the LHC energies. Thus, the chemical potential of the system is solely attributed to $\mu_{\rm B}$. We estimate the effect of rotation on rotational susceptibilities and their ratios in the VDWHRG model, which includes both attractive and repulsive interactions between the hadrons through $a$ and $b$ parameters. The model incorporates the contributions of all hadrons and resonances up to a mass cut-off of 2.25 GeV accessible in the Particle Data Group~\cite{PDG2016}. It is also worth noting that the thermodynamic pressure, higher-order fluctuations, and the correlations of conserved charges in a rotating medium may not be uniform, with components parallel and perpendicular to the angular momentum vector. However, for simplicity, the formalism here considers all the thermodynamic quantities and susceptibility observables to be isotropic in nature throughout the medium~\cite{Pradhan:2023rvf}. Furthermore, in our previous study~\cite{Pradhan:2023rvf}, the model is benchmarked against lQCD data at zero rotation for different thermodynamic quantities, baryon number, electric charge, and strangeness susceptibility. In this work, we extend the same model to estimate the rotational susceptibilities of hot and dense hadronic matter.\\

\begin{figure}
\includegraphics[scale=0.4]{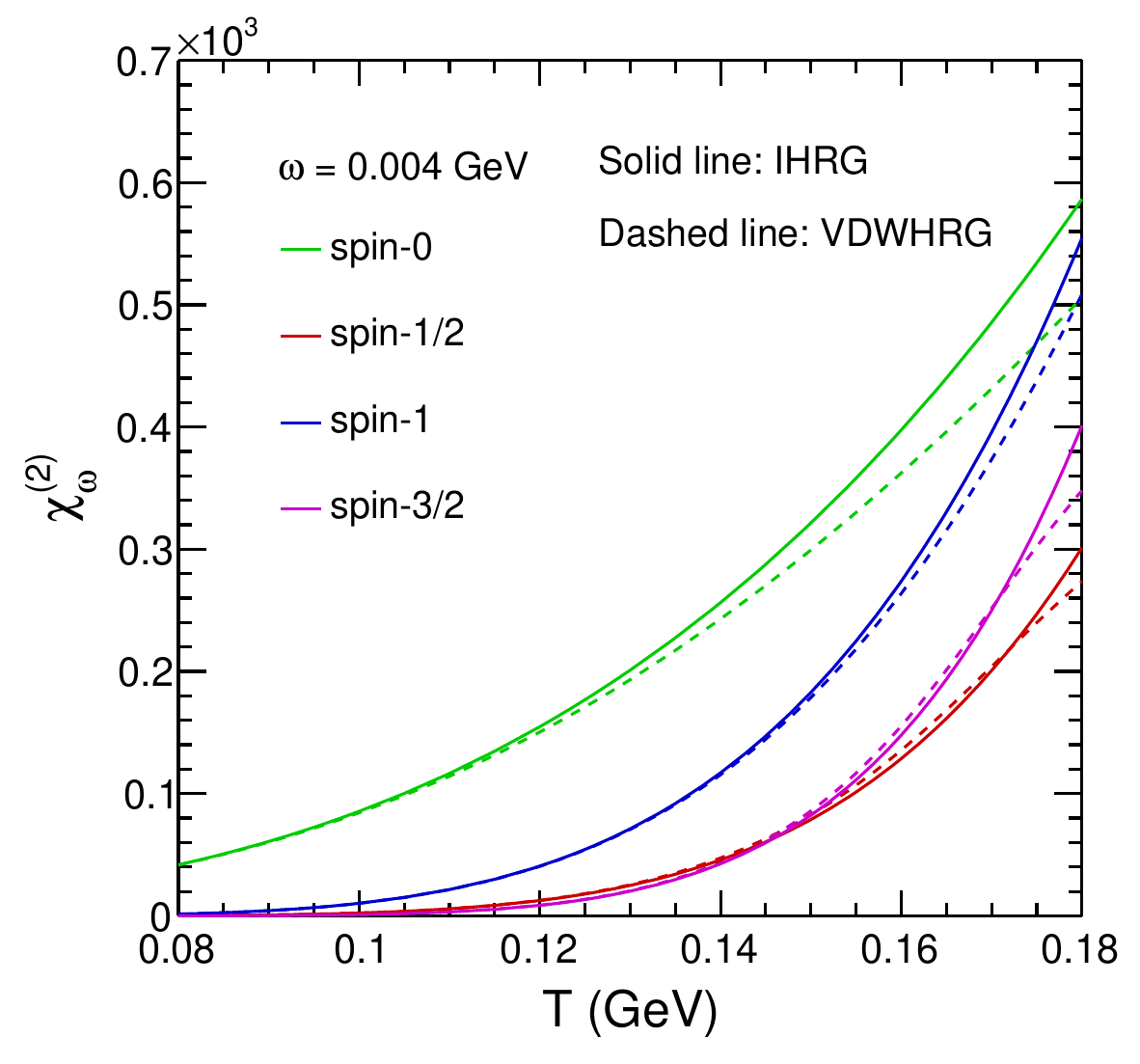}  
\includegraphics[scale=0.4]{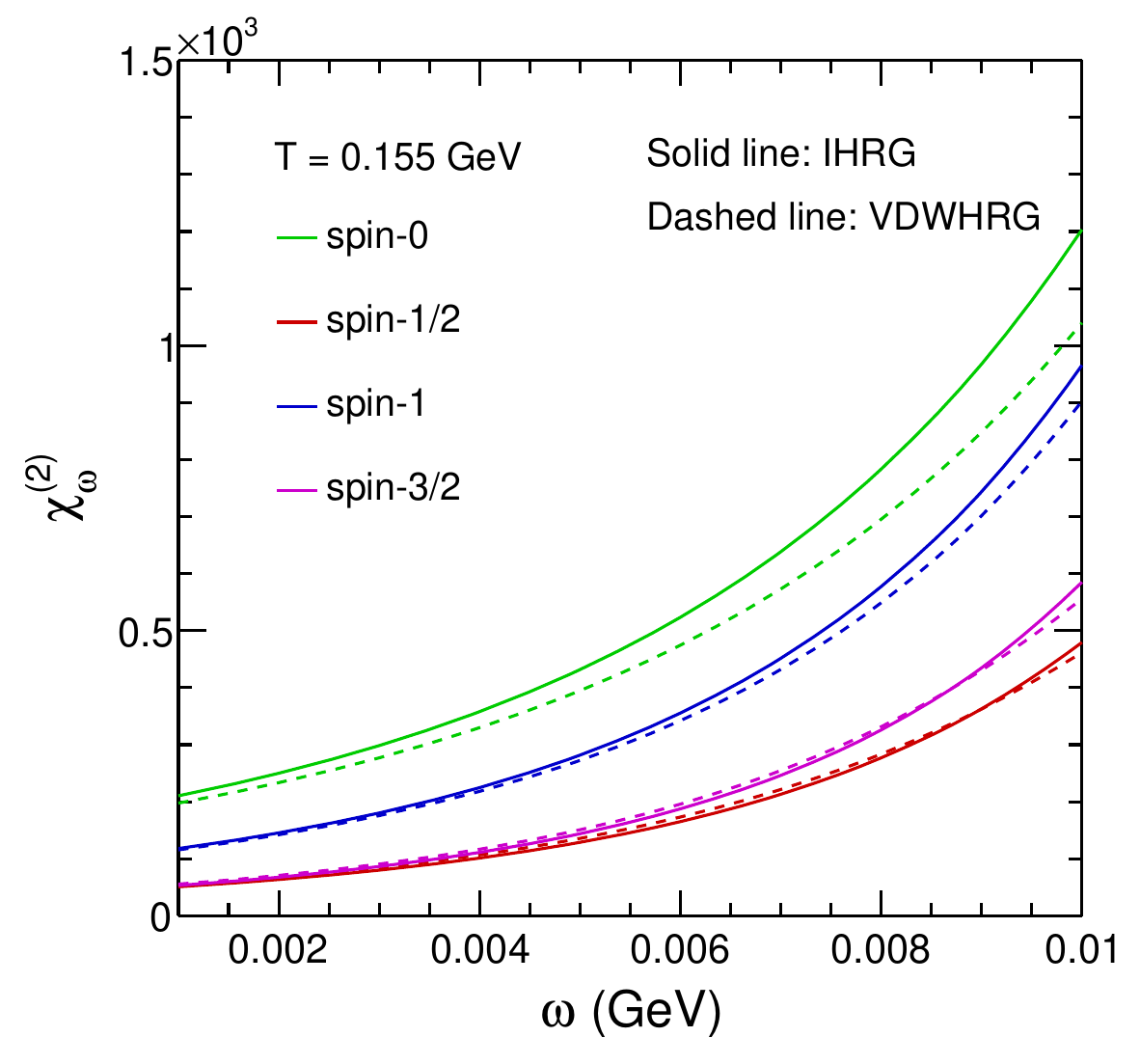}    
\caption{(Color online) The upper panel shows second-order rotational susceptibility $\chi^{(2)}_\omega$ as a functions of temperature for spin-0 (green), spin-1/2 (red), spin-1 (blue), and spin-3/2 (magenta) particles at $\omega$ = 0.004 GeV. The spin-0 particles have a substantial contribution to the second-order rotational susceptibility, followed by spin-1, spin-3/2, and spin-1/2 at higher temperatures. The lower panel displays $\chi^{(2)}_\omega$ as a function of $\omega$ for various spin particles at a temperature of T = 0.155 GeV, indicating that spin-dependent ordering of $\chi^{(2)}_\omega$ persists at higher values of $\omega$.}
\label{fig:spin}
\end{figure}

\begin{figure}
\includegraphics[scale=0.4]{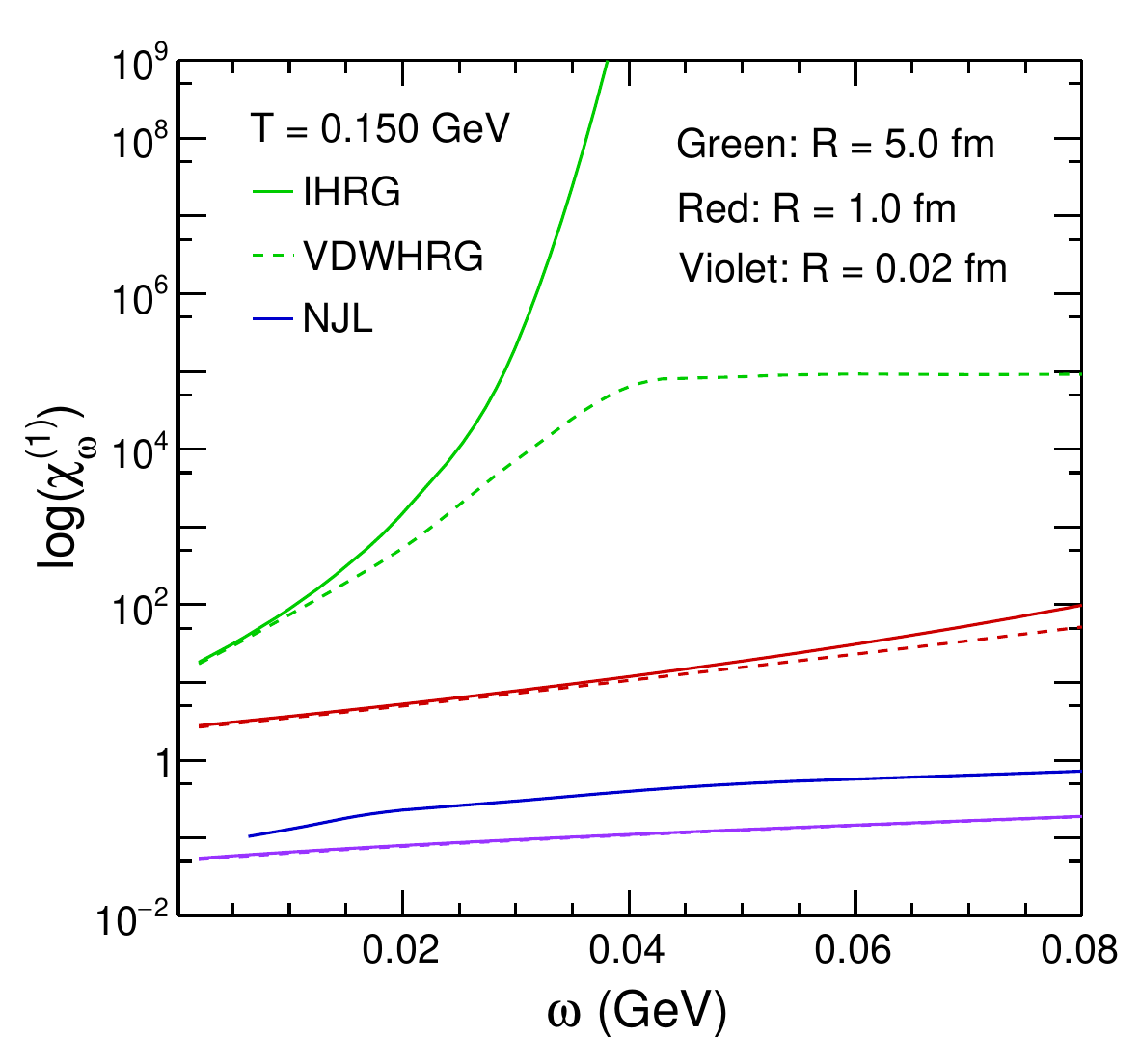}  
\includegraphics[scale=0.4]{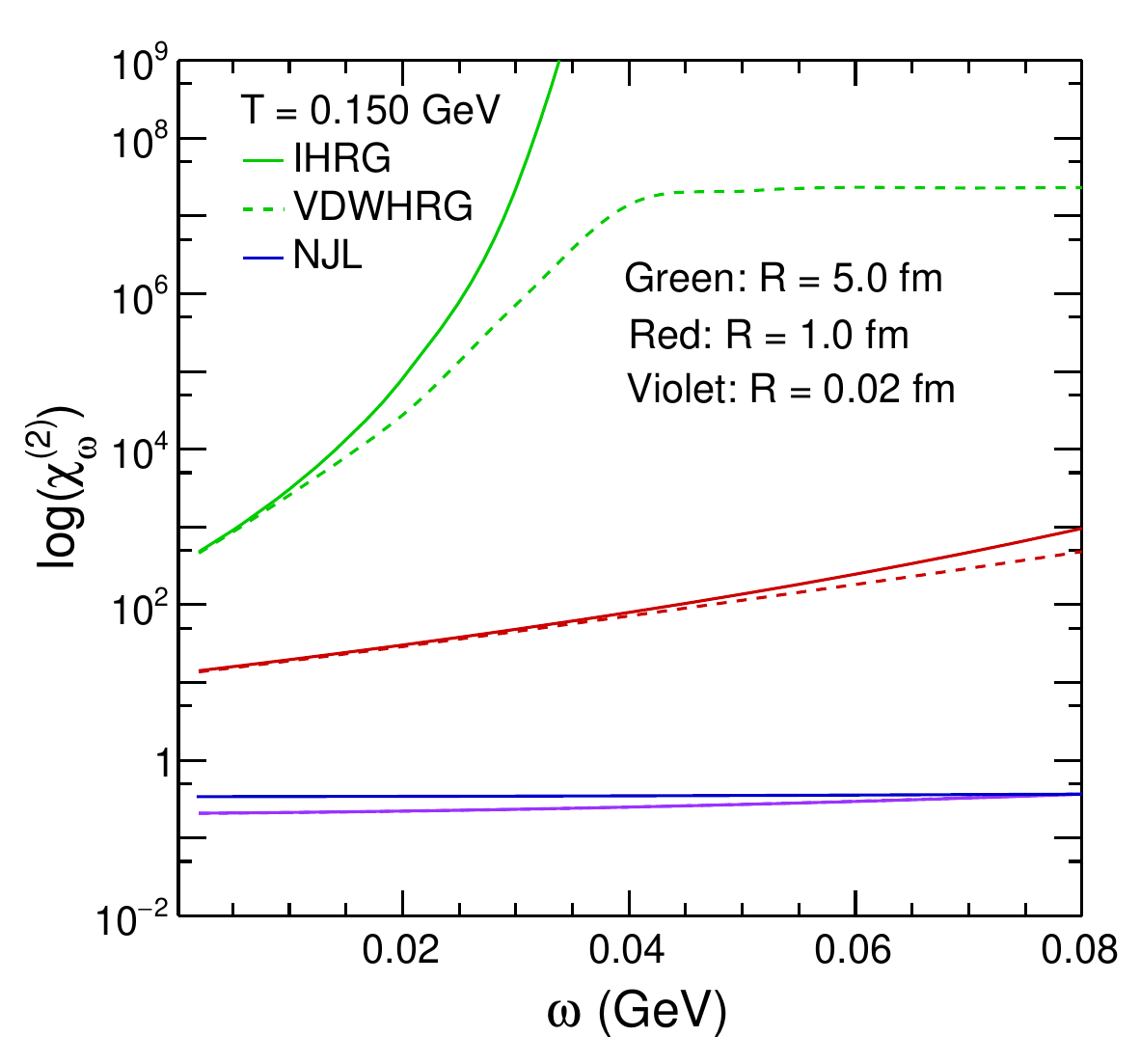}    
\caption{(Color online) The upper (lower) panel shows first (second) order rotational susceptibility $\chi^{(1)}_\omega$ ($\chi^{(2)}_\omega$) as a functions of rotation $\omega$ obtained in ideal HRG and VDWHRG model and compared with 3-flavour NJL model estimation at temperature T = 0.150 GeV, which considers $R$ = 0.02 fm ($\sim$ 0.1 GeV$^{-1}$)~\cite{Sun:2021hxo}. The comparison with the ideal HRG model serves as the baseline, while the NJL models suggest the qualitative agreement of our study as a function of rotation. Both the first- and second-order rotational susceptibilities show higher sensitivity to system size.}
\label{fig:NJL}
\end{figure}

Figure~\ref{fig:susvsC} shows the variation of $\chi^{(1)}_\omega$, $\chi^{(2)}_\omega$, $\chi^{(3)}_\omega$, $\chi^{(4)}_\omega$ with temperature at zero baryochemical potential for different values of $\omega$. The solid lines are the estimation from the ideal HRG model, whereas the dashed lines are those from the VDWHRG model. We will use this notation throughout the study. The comparison with ideal HRG provides a reference baseline against which interaction effects are explored in the VDWHRG model. In the present work, we examine four different values of rotation, such as $\omega$ = 0.002, 0.004, 0.006, and 0.008 GeV for our study. The considered values of $\omega$ in this study are close to following the recent experimental polarization measurement $\Lambda$ and $\bar{\Lambda}$ hyperons, $\omega$ $\simeq$ ($P_{\Lambda}$ + $P_{\bar{\Lambda}}$)$k_{B}$T/{$\hslash$} $\simeq$ (9 $\pm$ 1) $\times$ 10$^{21}$ sec$^{-1}$ $\simeq$ (0.004 - 0.008) GeV~\cite{STAR:2017ckg}. We find that the magnitude of rotational susceptibilities increases monotonically from the first to fourth order of susceptibilities for all values of temperature and rotation. In addition, at a given temperature, a higher $\omega$ gives a higher susceptibility, implying that the rotation enhances the different orders of rotational susceptibilities or the so-called spin polarization.  A similar finding for first-order susceptibility is reported in the three-flavor NJL model~\cite{Sun:2021hxo}. Furthermore, considering the interactions among the hadrons, the rotational susceptibility value is suppressed compared to that in the non-interacting case. \\ 
\begin{figure*}
\centering
\includegraphics[scale=0.29]{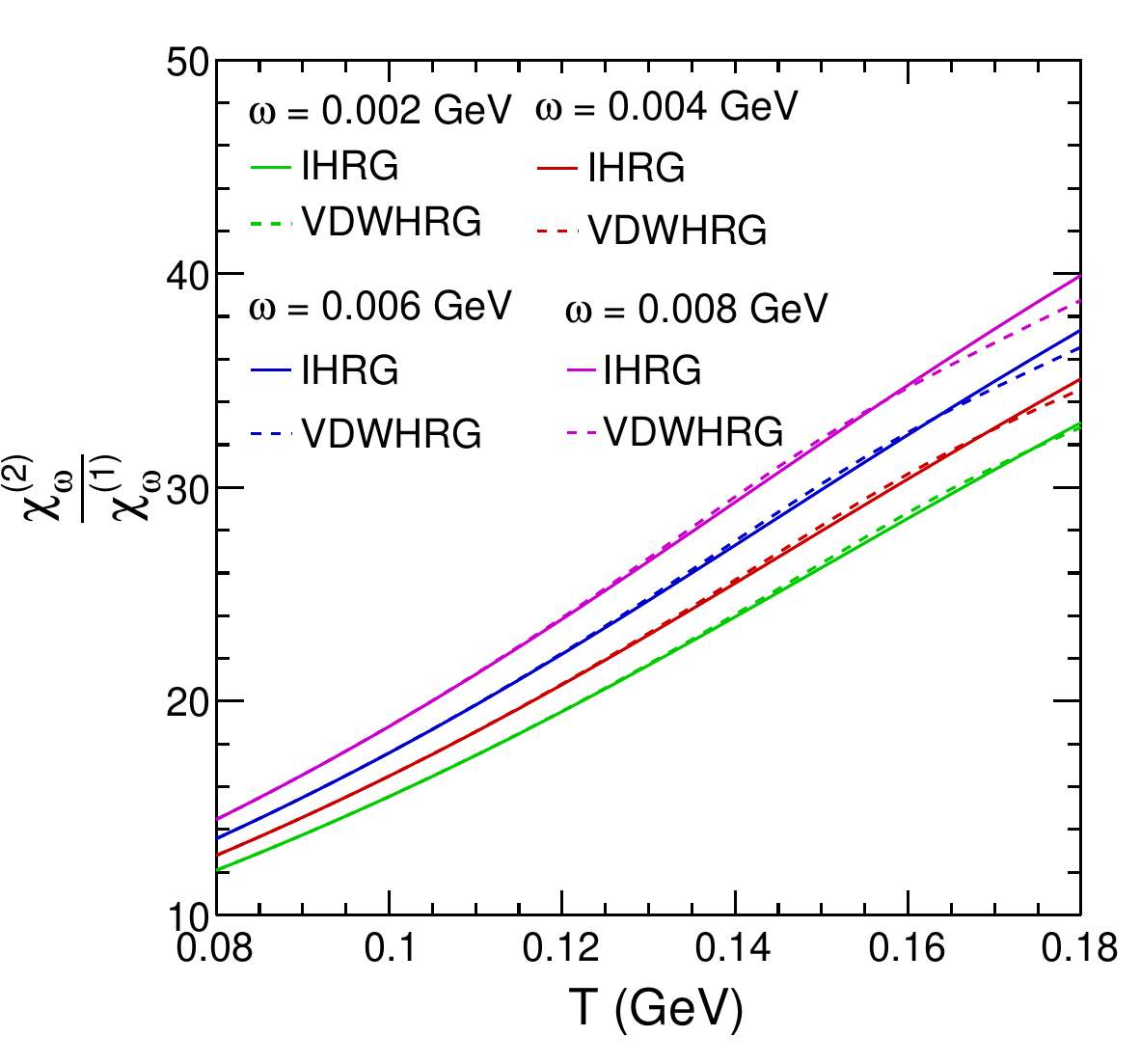}
\includegraphics[scale=0.29]{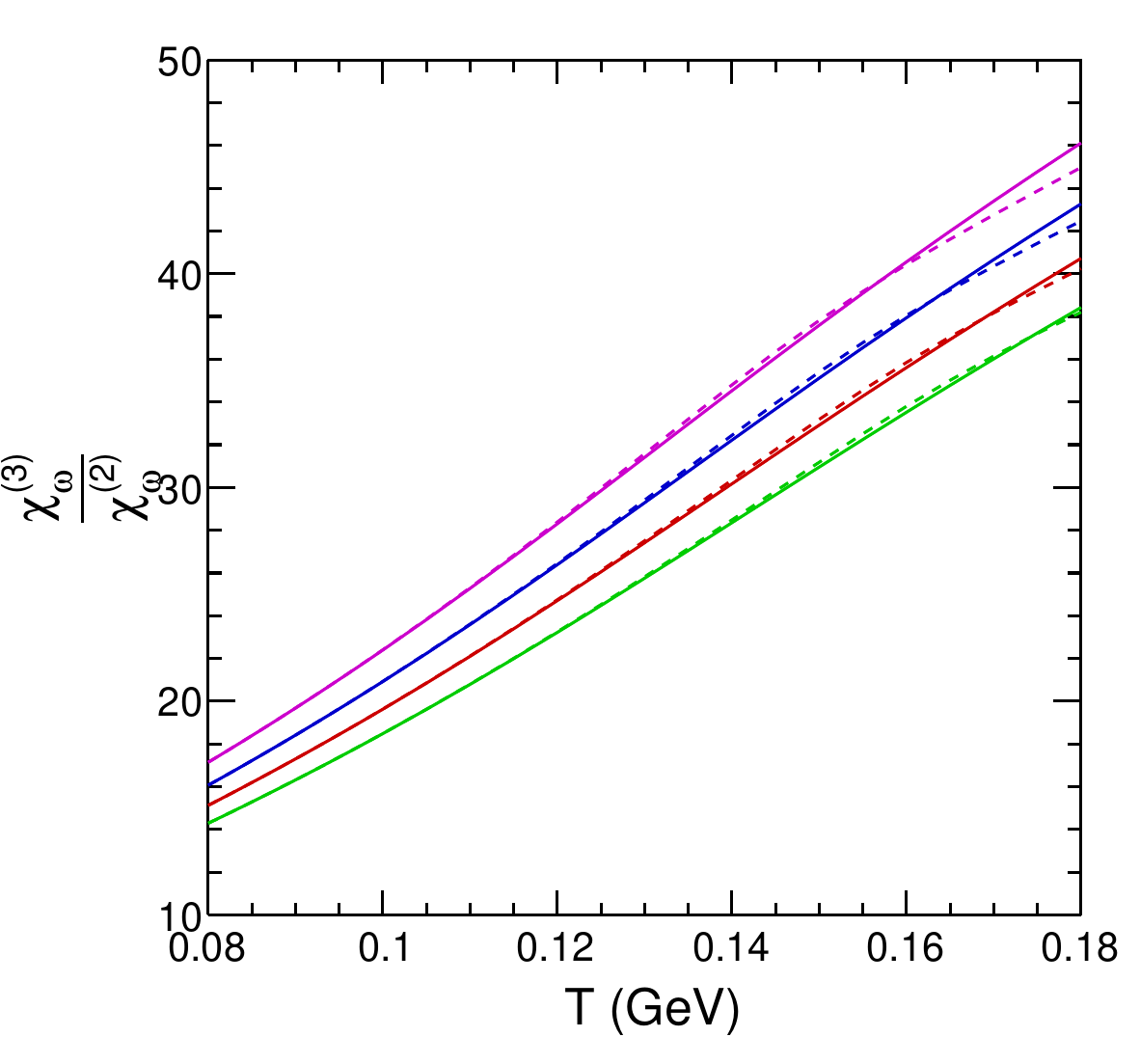}
\includegraphics[scale=0.29]{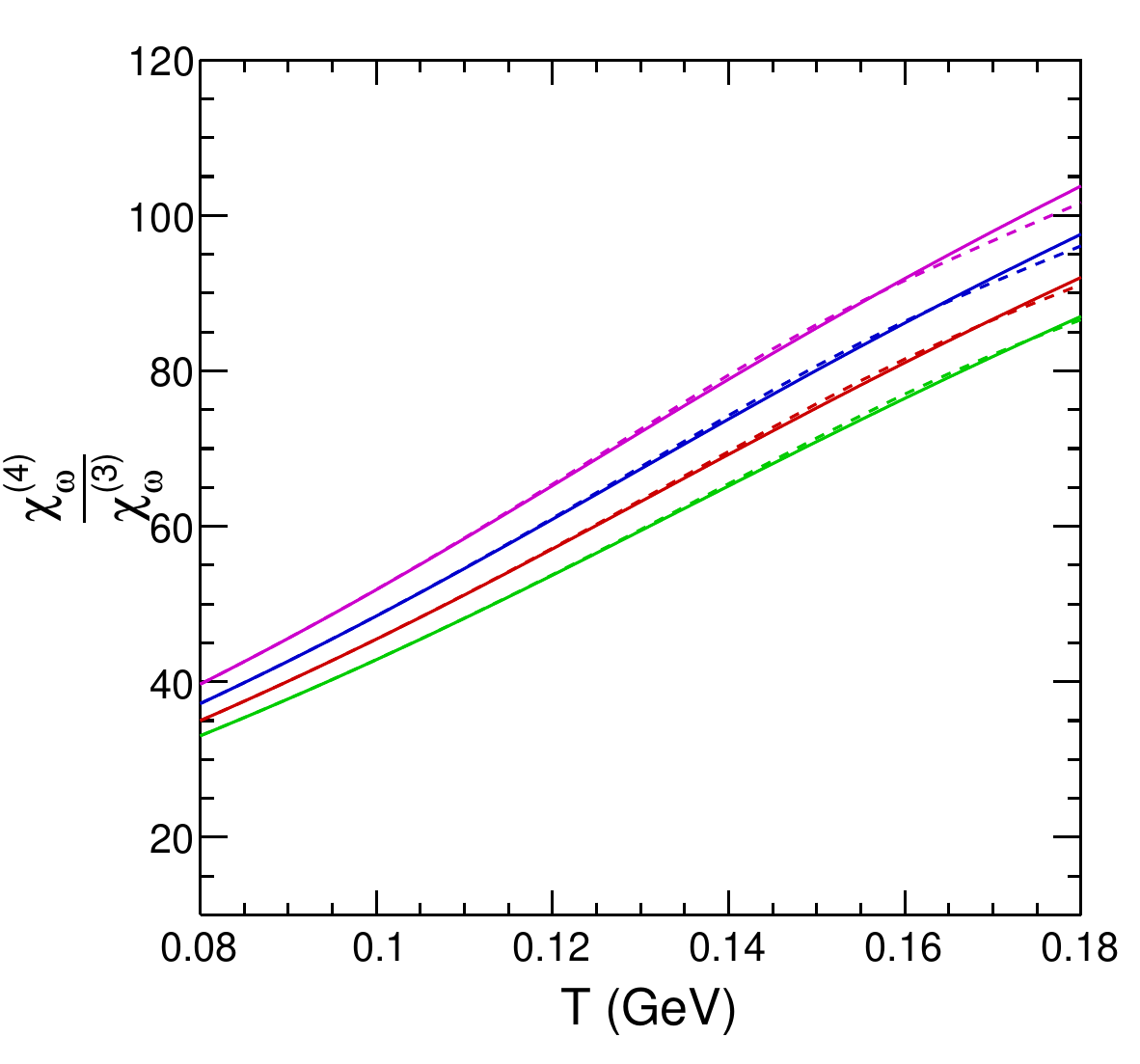}  
\includegraphics[scale=0.29]{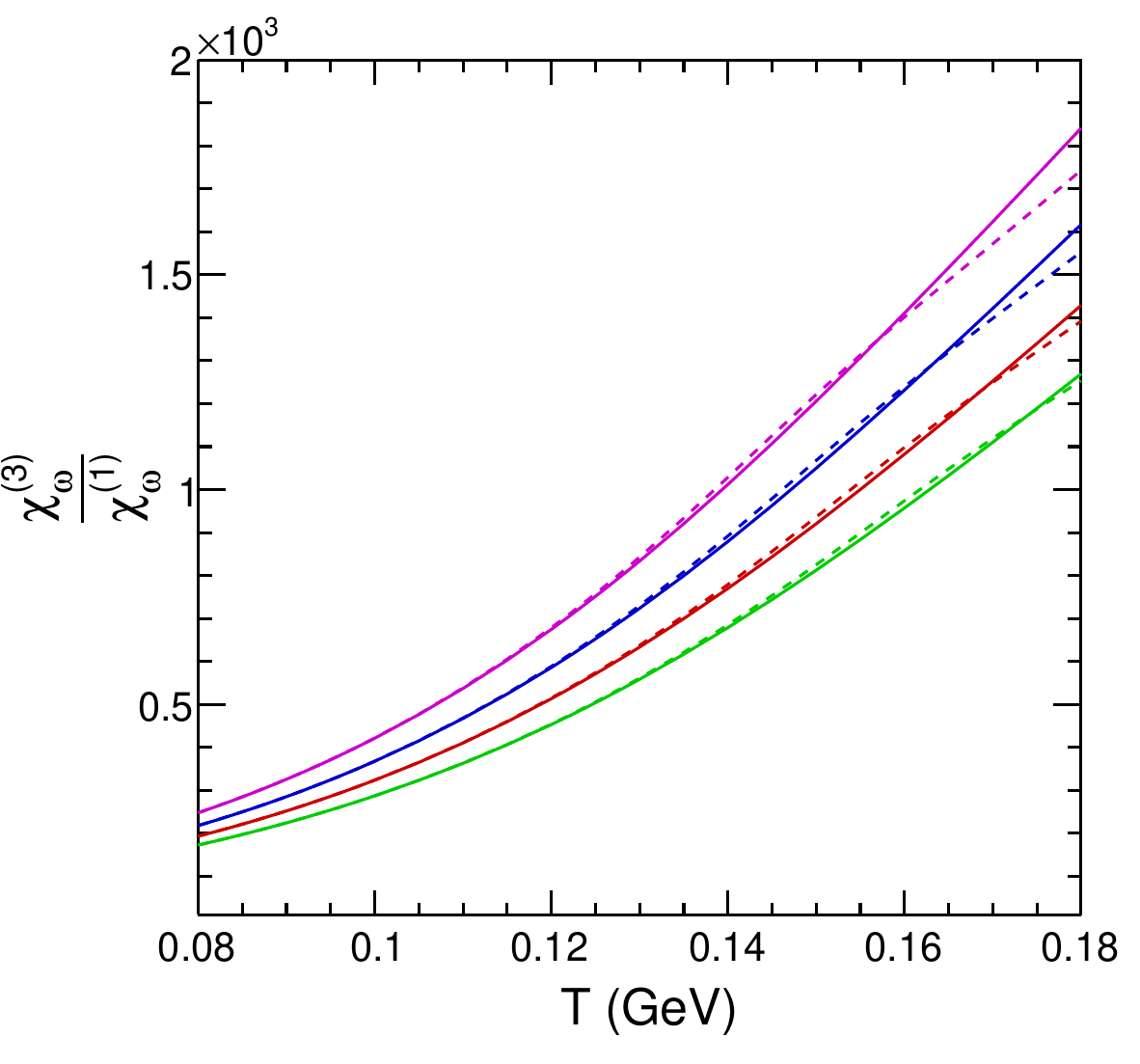}
\includegraphics[scale=0.29]{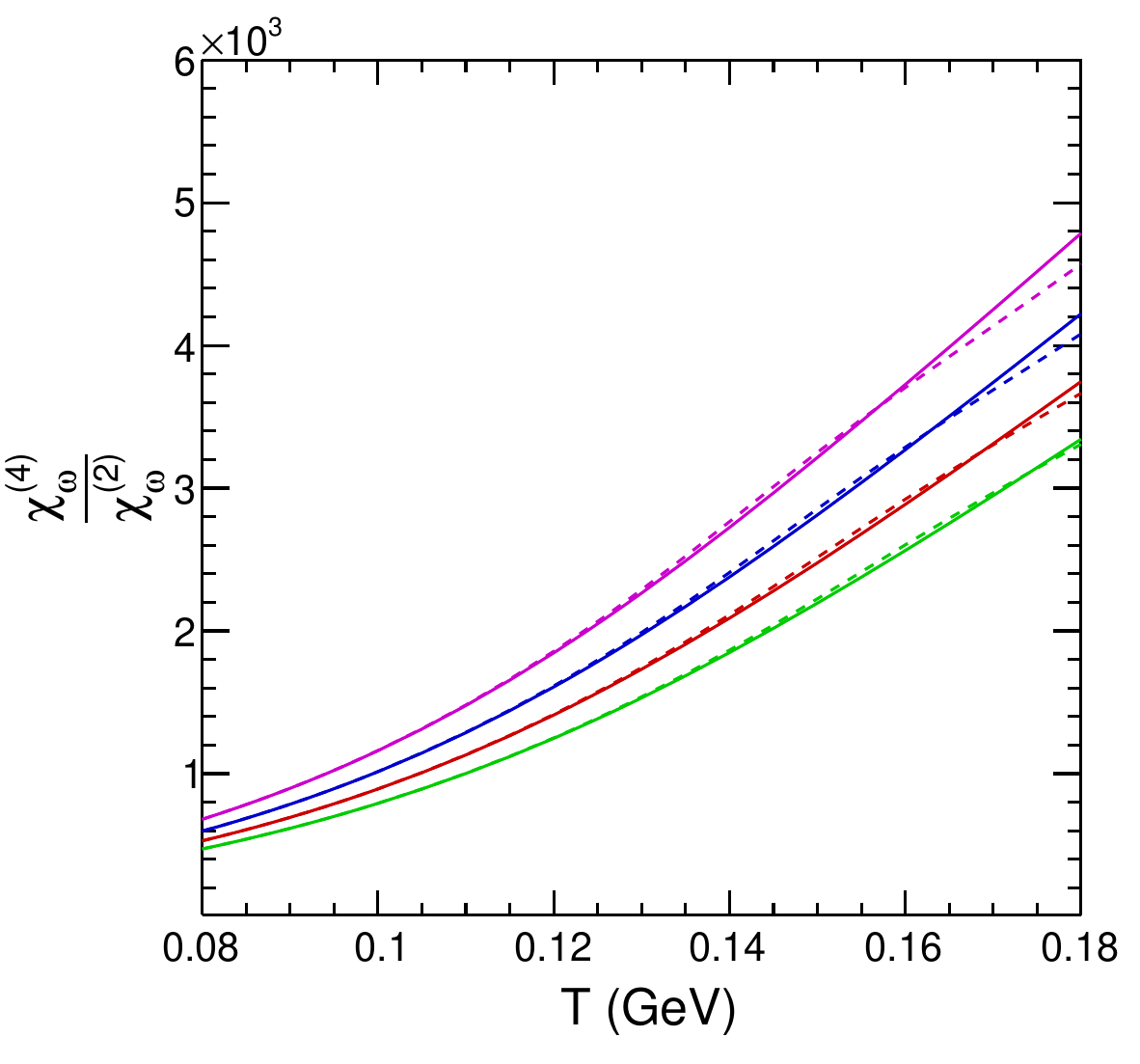}
\caption{(Color online) The variation (from left to right and downwards) of $\chi^{(2)}_\omega$/$\chi^{(1)}_\omega$, $\chi^{(3)}_\omega$/$\chi^{(2)}_\omega$, $\chi^{(4)}_\omega$/$\chi^{(3)}_\omega$, $\chi^{(3)}_\omega$/$\chi^{(1)}_\omega$, $\chi^{(4)}_\omega$/$\chi^{(2)}_\omega$ ratios as a functions of temperature at zero baryochemical potential ($\mu_{\rm B}$ = 0 GeV), for $\omega$ = 0.002 GeV (green), $\omega$ = 0.004 GeV (red), $\omega$ = 0.006 GeV (blue), and $\omega$ = 0.008 GeV (magenta) in ideal HRG (solid line) and VDWHRG (dotted line) model. A distinct ordering pattern in the even and odd powers of rotational susceptibility ratios is observed, and its sensitivity to temperature and rotation is examined in both the ideal HRG and VDWHRG models.}
\label{fig:susratiosvsC}
\end{figure*}

The first-order derivative of pressure (or free energy) with respect to rotation is called the angular momentum density, which is similar to the number density defined as the first-order derivative of pressure with respect to chemical potential. In general, the spin density contains crucial information about the system, such as the spin polarization of hadrons induced by rotation (called rotational polarization). The angular momentum, as defined in Eq.~\ref{eq4}, and the moment of inertia (ratio of the angular momentum to medium rotation) can also be calculated from the first-order derivative of free energy with respect to rotation. From Fig.~\ref{fig:susvsC}, it can be inferred that the moment of inertia of the hadronic matter is positive. However, a negative moment of inertia is obtained in lQCD calculation for a rotating quark-gluon plasma~\cite{Braguta:2023tqz, Braguta:2023yjn, Braguta:2023iyx}. The negative moment of inertia was argued to emerge from the negativity of a gluonic Barnett eﬀect, which implies a negative coupling of the gluon spin polarization with vorticity. The difference in the moment of inertia in the hadronic matter and QGP phases is due to the different degrees of freedom and different equations of state used in the two systems.\\

Moreover, studying the first-order and higher-order susceptibilities due to rotation and understanding the role of spin to describe the hot hadronic/QCD matter evolution is an important topic of research in relativistic heavy-ion collisions. With this motivation, we have explicitly plotted $\chi^{(2)}_{\omega}$ for spin-0, spin-1/2, spin-1, and spin-3/2 particles in a hadron gas to understand the effect of spin on the rotational susceptibility observable. The upper and lower panel of Fig.~\ref{fig:spin} shows the variation of $\chi^{(2)}_{\omega}$ as a function of temperature and $\omega$, respectively, for spin-0, spin-1/2, and spin-1, spin-3/2 hadrons for a rigidly rotating system. We observe that at lower temperatures ($\sim (0.080-0.130)$ GeV), $\chi^{(2)}_{\omega}$ is found to have negligible contribution from spin-1/2 and spin-1, spin-3/2 particles, while spin-0 particles have substantial contribution. However, at high temperatures, all the baryons start populating the hadronic matter and contribute to the thermodynamic susceptibility significantly. At high temperature, the spin-0 particles (mainly dominated by pions ($\pi^{\pm}$, $\pi^{0}$)) have the maximum contribution to $\chi^{(2)} _{\omega}$, followed by spin-1 ($\rho^{\pm}$, $\rho^{0}$), spin-3/2 ($\Delta^{++}$, $\Delta^{+}$, $\Delta^{0}$, $\Delta^{-}$), and spin-1/2 [protons ($p$), neutrons ($n$)]. The lower panel of Fig.~\ref{fig:spin} shows the second-order rotational susceptibility $\chi^{(2)}_{\omega}$ obtained at T = 0.155 GeV, and increases as a function of rotation for all kinds of particles. Moreover, a system of hadron gas with only spin-0 particles has a more pronounced response to rotation and the corresponding susceptibility $\chi^{(2)}_{\omega}$, followed by  spin-1, spin-3/2, and spin-1/2 particles. The van der Waals interactions among the hadrons lower the $\chi^{(2)}_{\omega}$ as compared to the no interaction case.\\

\begin{figure*}
\includegraphics[scale=0.37]{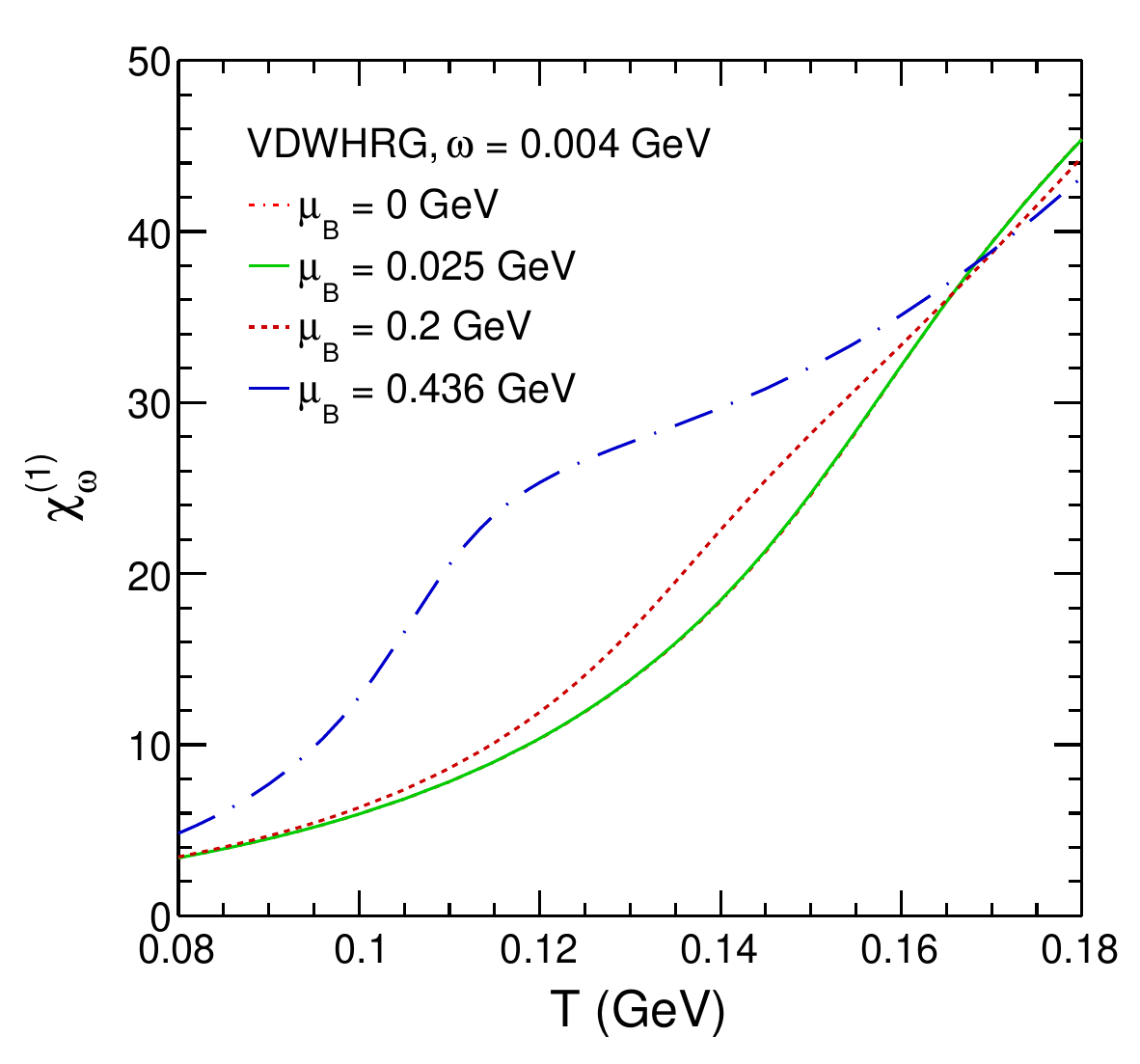}
\includegraphics[scale=0.37]{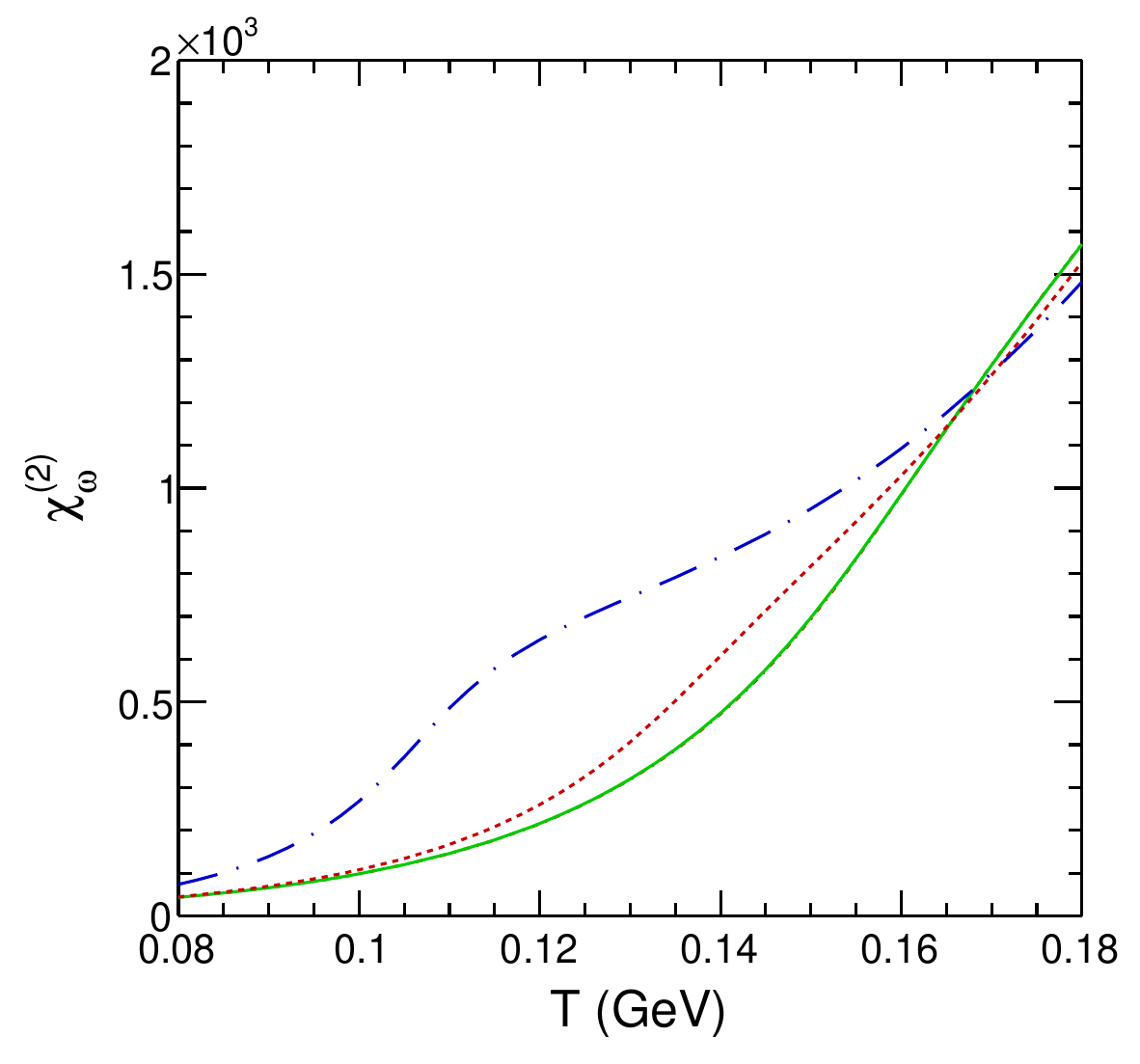}
\includegraphics[scale=0.37]{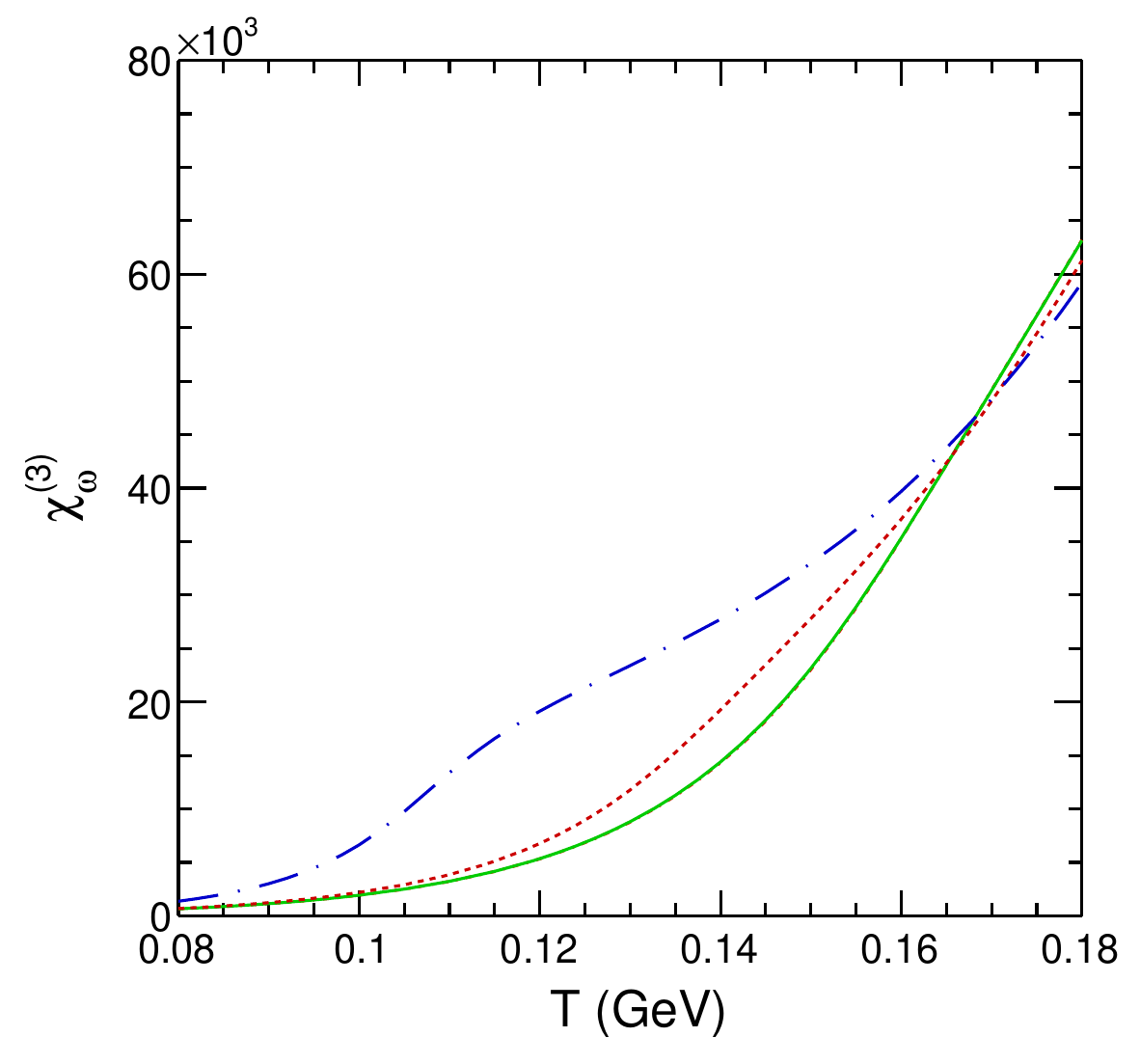}
\includegraphics[scale=0.37]{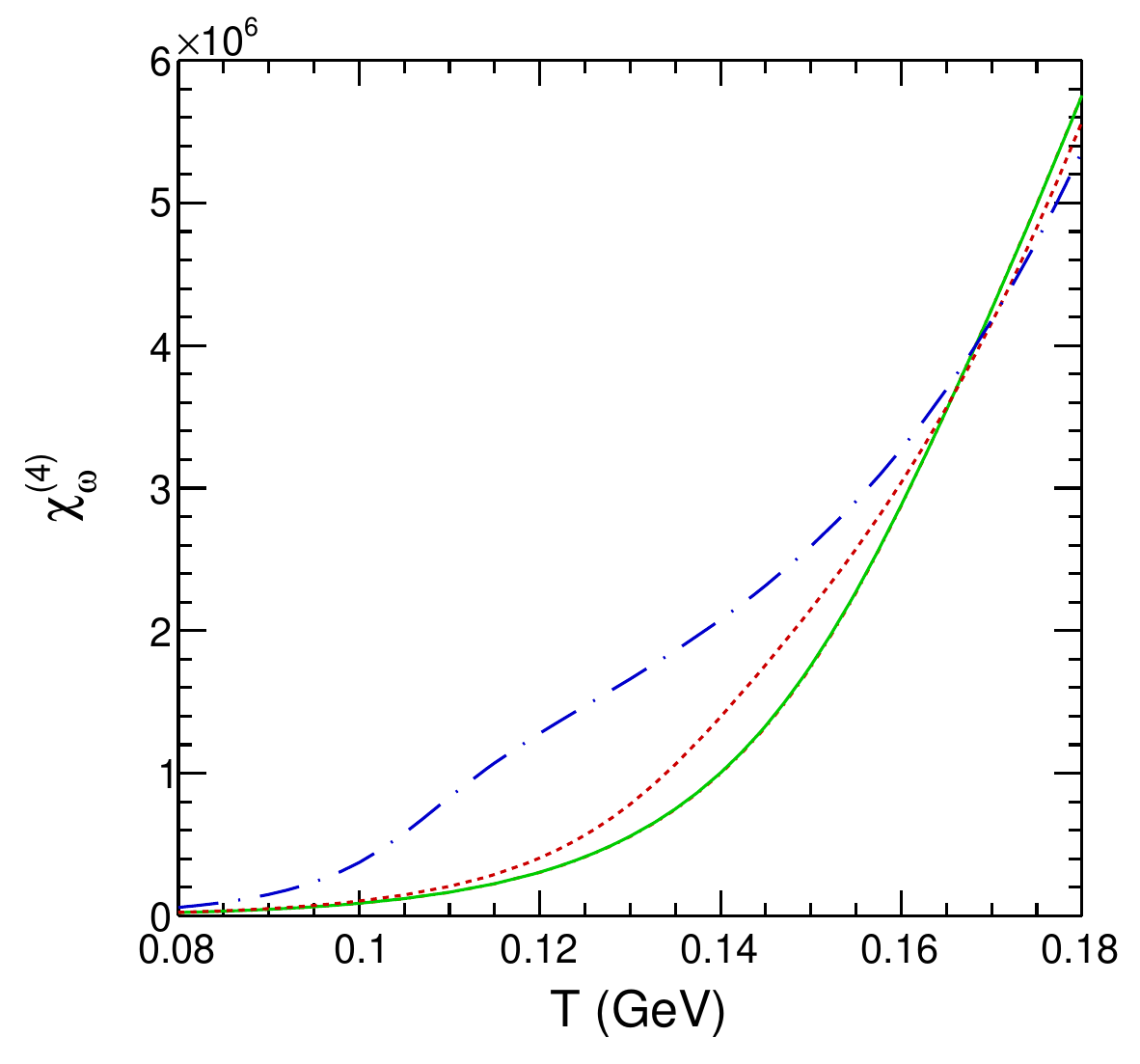}  
\caption{(Color online) The variation (from left to right and downwards) of $\chi^{(1)}_\omega$, $\chi^{(2)}_\omega$, $\chi^{(3)}_\omega$, $\chi^{(4)}_\omega$ as a functions of temperature at different baryochemical potentials for $\omega$ = 0.004 GeV in VDWHRG model. The non-monotonic structure (\enquote{bump}) seen in rotational susceptibilities at low temperatures and high baryon densities is probed via rotation and mainly stems from the van der Waals equation of state, which exhibits the nuclear liquid-gas phase transition~\cite{Vovchenko:2015vxa, Vovchenko:2015pya}.}
\label{fig:susvsmu}
\end{figure*}

Figure~\ref{fig:NJL} shows the comparison of first-order (upper panel) and second-order (lower panel) rotational susceptibility as a function of rotation obtained from the ideal HRG, VDWHRG, and the three-flavor NJL model calculation at $T$ = 0.150 GeV. We found that our estimation qualitatively agrees with the NJL model calculation at $R$ = 0.02 fm ($\sim$ 0.1 GeV$^{-1}$)~\cite{Sun:2021hxo}, where the same value of $R$ is used following Ref.~\cite{Jiang:2016wvv}. Further, the system size dependence of first- and second-order rotational susceptibility is shown in Fig.~\ref{fig:NJL} for two different values of $R$ = 1 fm and 5 fm. The values of first- and second-order rotational susceptibility increase with an increase in system size as well as rotation. The VDWHRG model predicts a saturation behavior at high values of $\omega$, for $R$ = 5 fm.  This saturation can be attributed to two main factors. First, in the VDWHRG model, repulsive interactions among hadrons restrict the increase in energy density and response functions at high angular velocities, which is the reason for it to be lower than the ideal HRG value. Second, causality imposes an upper bound on the tangential velocity, $v = \omega R$, which approaches the speed of light for large $\omega = 0.04 ~\rm$ GeV and $R = 5$ fm. As this limit is approached, the system can no longer respond linearly to increasing rotation, resulting in the observed saturation behavior. It should be emphasized that the NJL model comparison offers a qualitative check for rotational susceptibility at small $\omega$ within different effective descriptions of strongly interacting matter. The HRG and NJL models describe different physical regimes, corresponding to hadronic and quark degrees of freedom, respectively. The comparison does not demonstrate (i) continuity across the hadron–quark transition, (ii) sensitivity to chiral symmetry restoration, and (iii) a link to QCD critical dynamics.


\begin{figure*}
\includegraphics[scale=0.29]{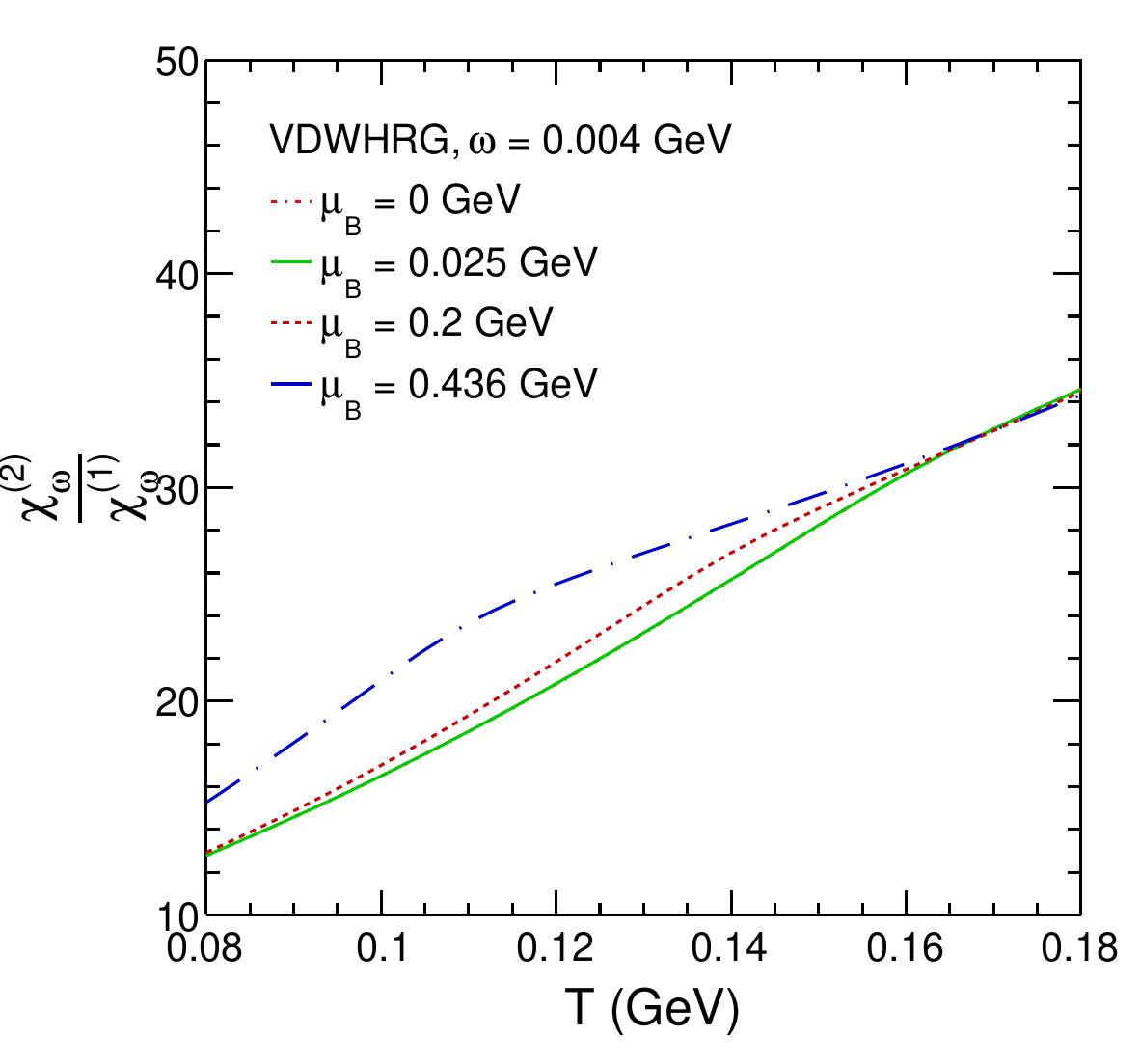}
\includegraphics[scale=0.29]{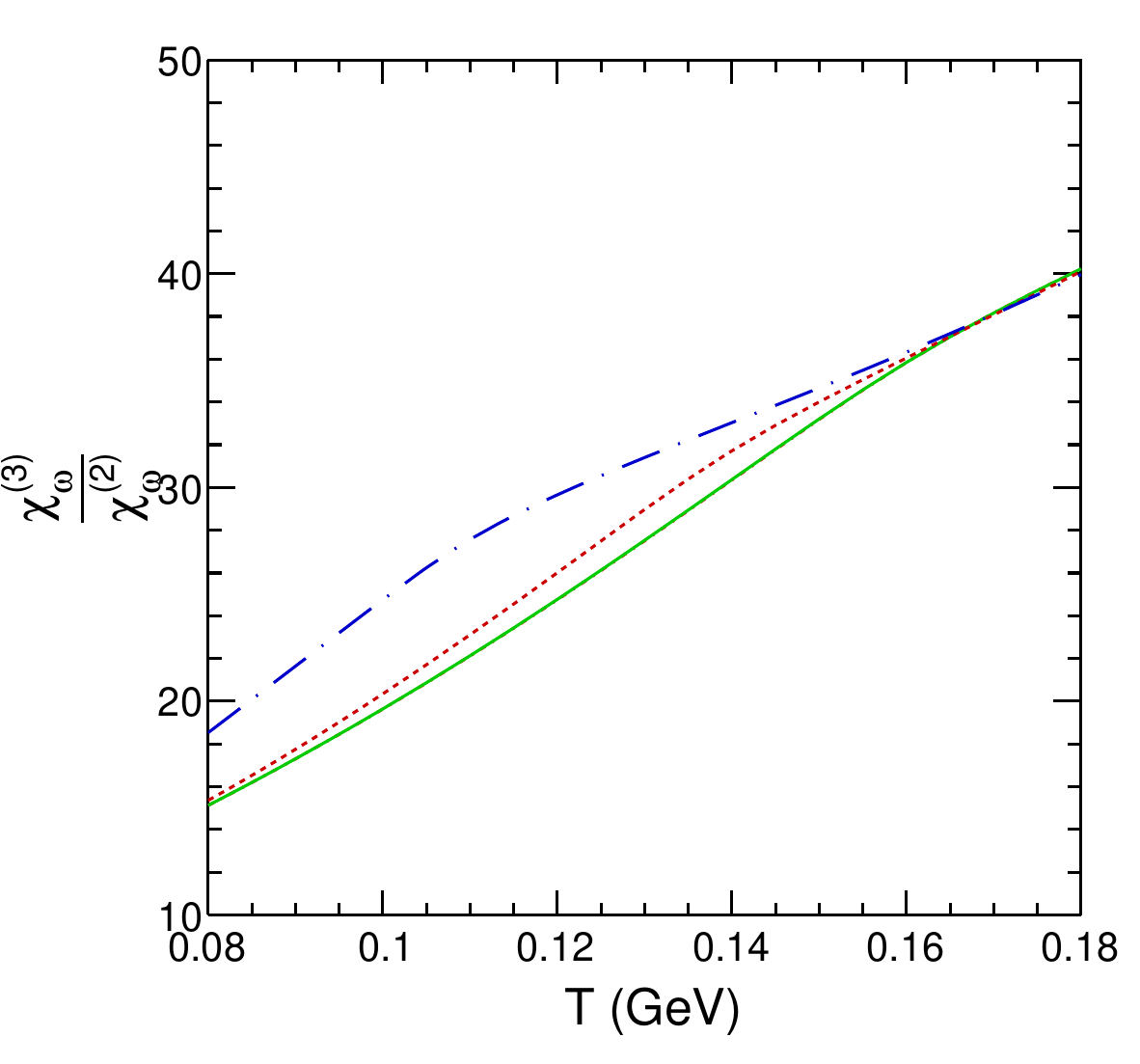}
\includegraphics[scale=0.29]{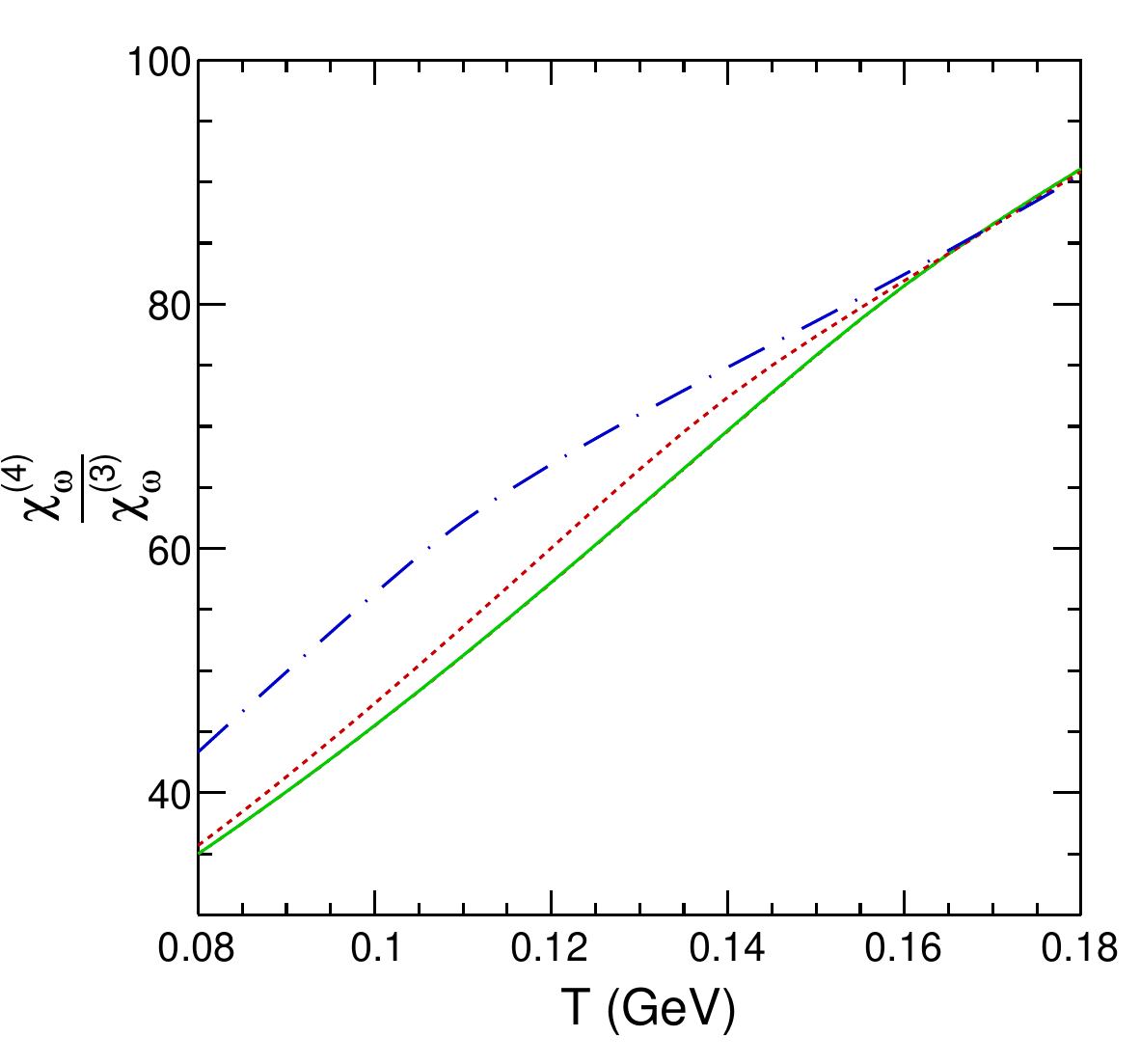}   
\includegraphics[scale=0.29]{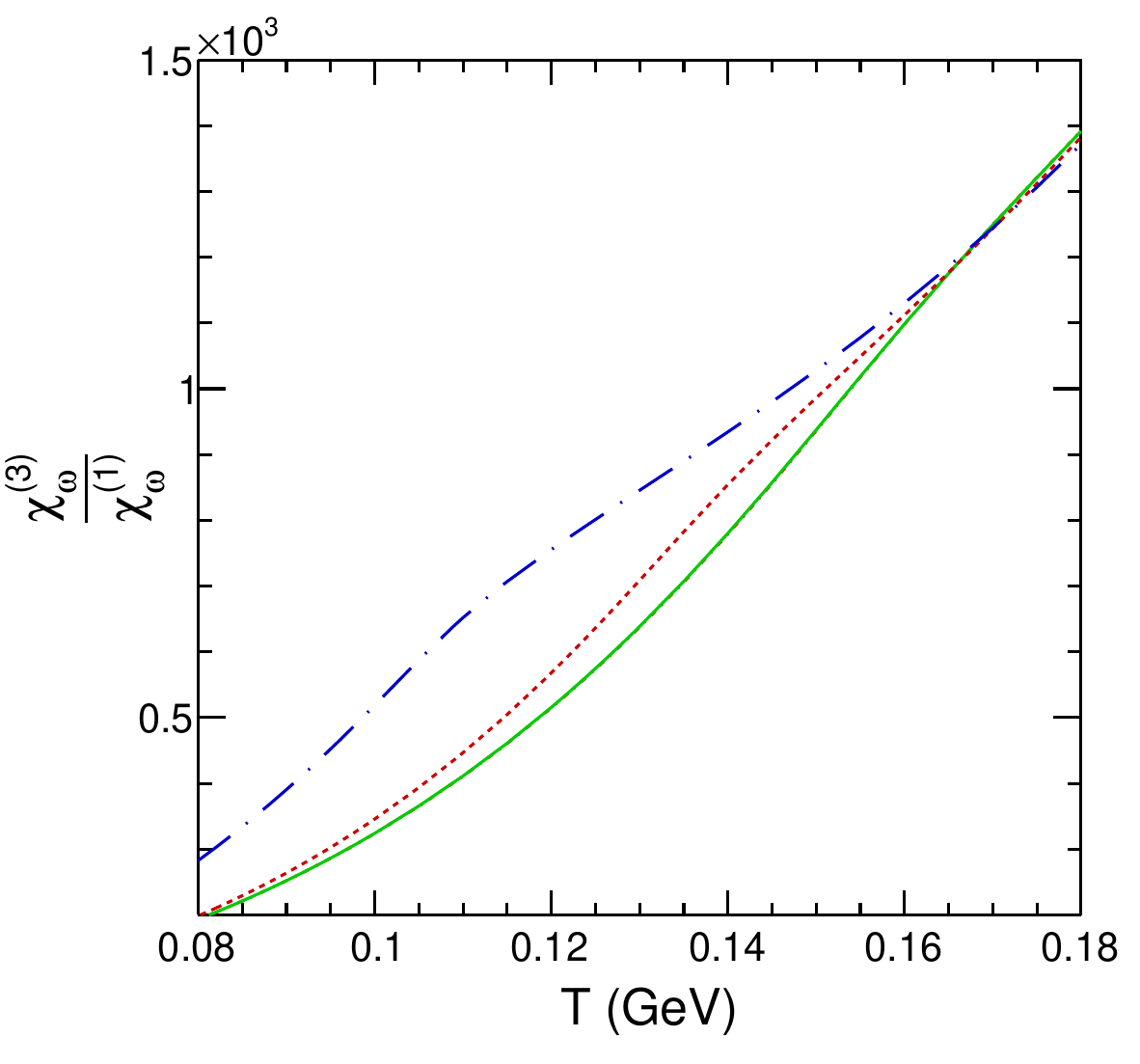}
\includegraphics[scale=0.29]{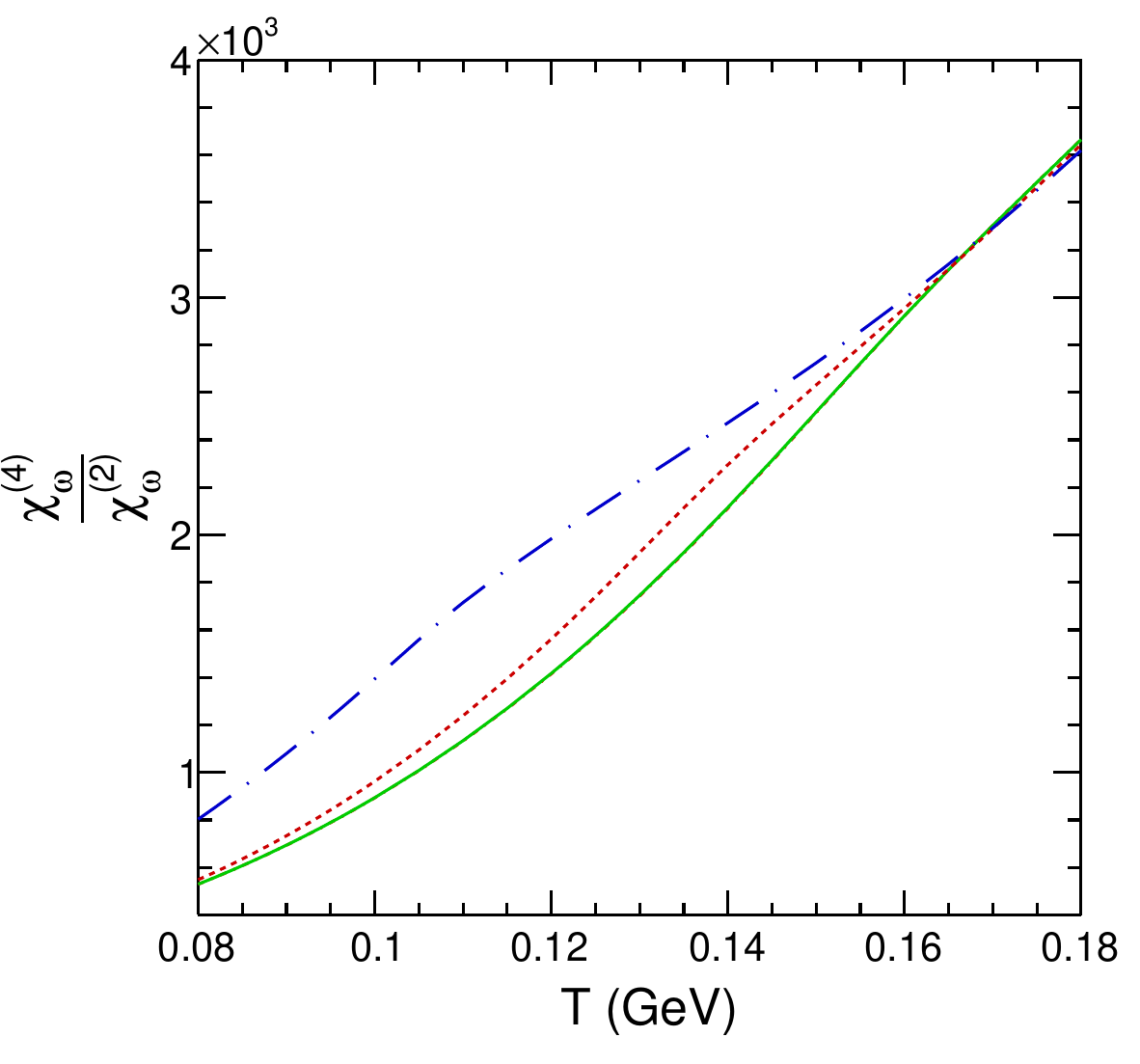}
\caption{(Color online)  The variation (from left to right and downwards) of $\chi^{(2)}_\omega$/$\chi^{(1)}_\omega$, $\chi^{(3)}_\omega$/$\chi^{(2)}_\omega$, $\chi^{(4)}_\omega$/$\chi^{(3)}_\omega$, $\chi^{(3)}_\omega$/$\chi^{(1)}_\omega$, $\chi^{(4)}_\omega$/$\chi^{(2)}_\omega$ ratios as a functions of temperature at different baryochemical potentials for $\omega$ = 0.004 GeV in VDWHRG model. Susceptibility ratios behave similarly to susceptibility as a function of baryon chemical potential and temperatures, but the non-monotonic structure at $\mu_{B}$ = 0.436 GeV appears less pronounced in the ratio.}
\label{fig:susratiovsmu}
\end{figure*}

Figure~\ref{fig:susratiosvsC} illustrates the variation of $\chi^{(2)}_\omega$/$\chi^{(1)}_\omega$, $\chi^{(3)}_\omega$/$\chi^{(2)}_\omega$, $\chi^{(4)}_\omega$/$\chi^{(3)}_\omega$, $\chi^{(3)}_\omega$/$\chi^{(1)}_\omega$, $\chi^{(4)}_\omega$/$\chi^{(2)}_\omega$ ratios as a functions of temperature $T$ and rotation $\omega$ at zero baryon chemical potential. It is observed that all these ratios increase with an increase in $T$ and $\omega$ for both ideal HRG and VDWHRG models, although the slope of the curves varies as a function of $T$ and $\omega$. Interestingly, we found that the interactions among the hadrons play a crucial role in the rotational susceptibility ratios as functions of temperature. At intermediate temperature ranges ($\sim (0.120-0.160)$ GeV), the rotational susceptibility ratios are higher for the VDWHRG model; however, at higher temperatures ($\sim (0.160-0.180)$ GeV) and higher rotation ($\sim (0.006-0.008)$ GeV), the interaction suppresses the rotational susceptibility ratios. 


\begin{figure*}
\centering
\includegraphics[scale=0.37]{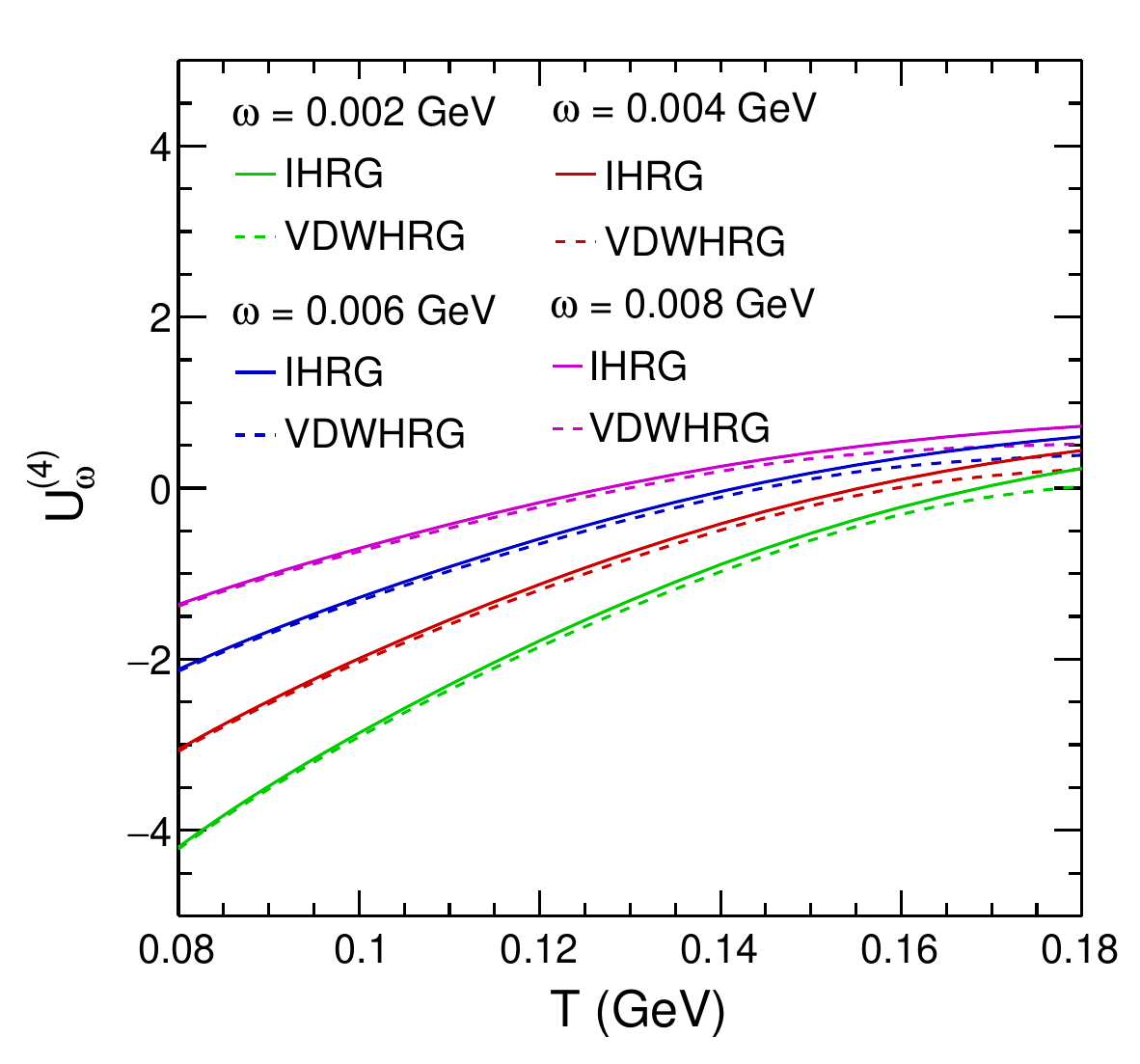}
\includegraphics[scale=0.37]{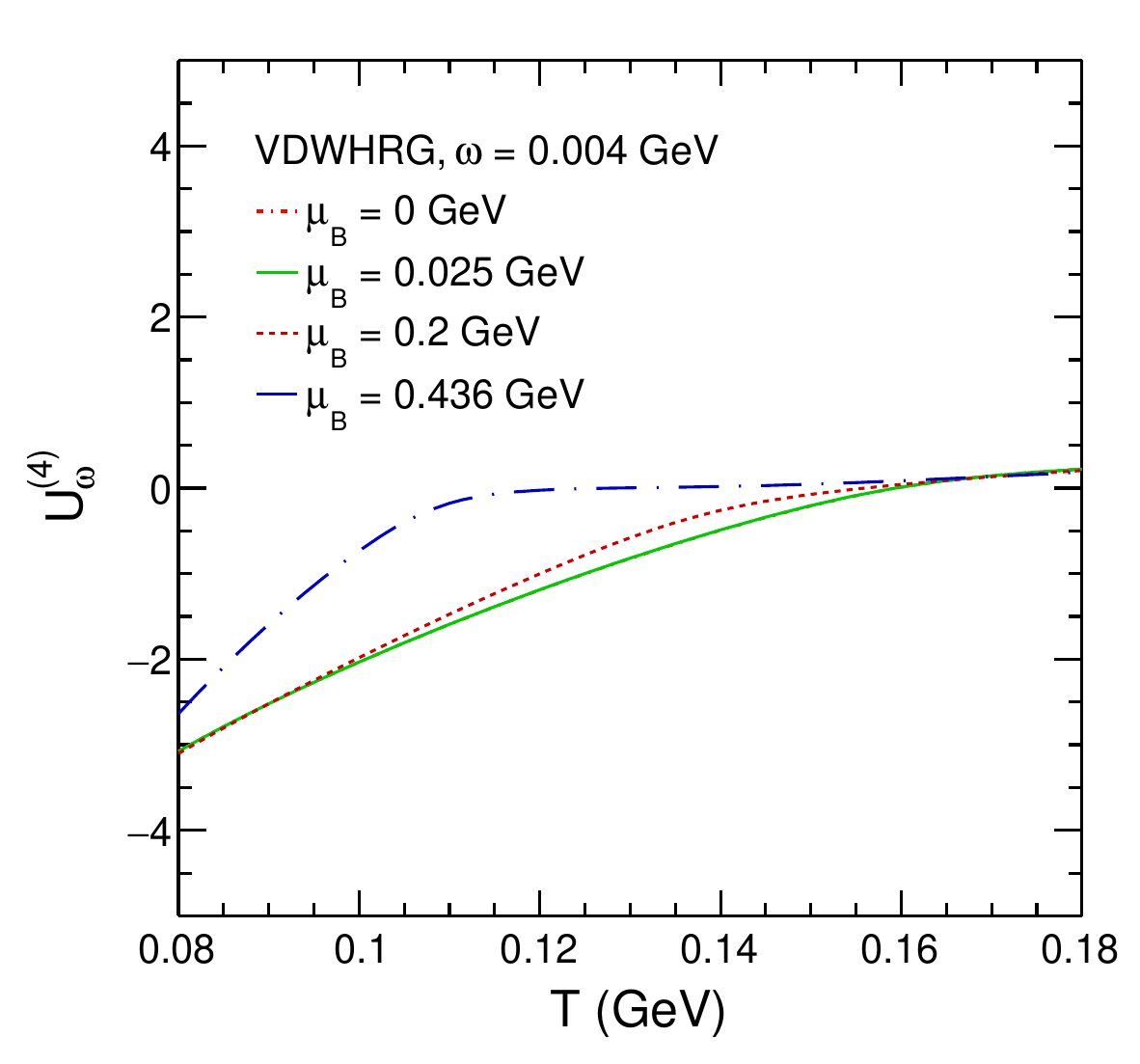}
\caption{The left panel shows the Binder cumulant as a function of temperature at zero baryochemical potential ($\mu_{\rm B}$ = 0 GeV), for $\omega$ = 0.002 GeV (green), $\omega$ = 0.004 GeV (red), $\omega$ = 0.006 GeV (blue), and $\omega$ = 0.008 GeV (magenta). The right panel shows Binder cumulant as a function of as a functions of temperature at different baryochemical potentials for $\omega$ = 0.004 GeV in the VDWHRG model.
}
\label{U4}
\end{figure*}

\begin{figure*}
\includegraphics[scale=0.29]{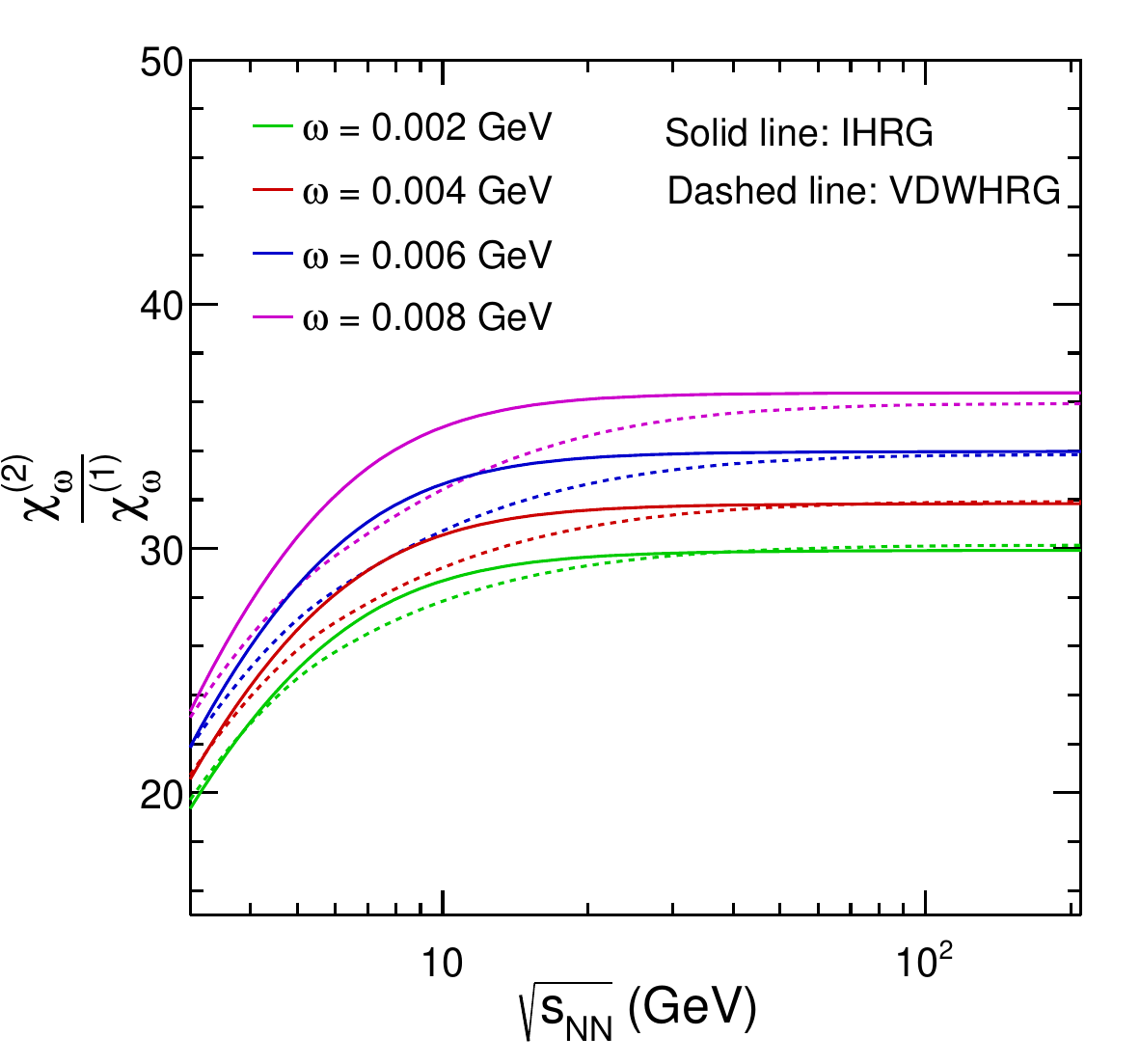}
\includegraphics[scale=0.29]{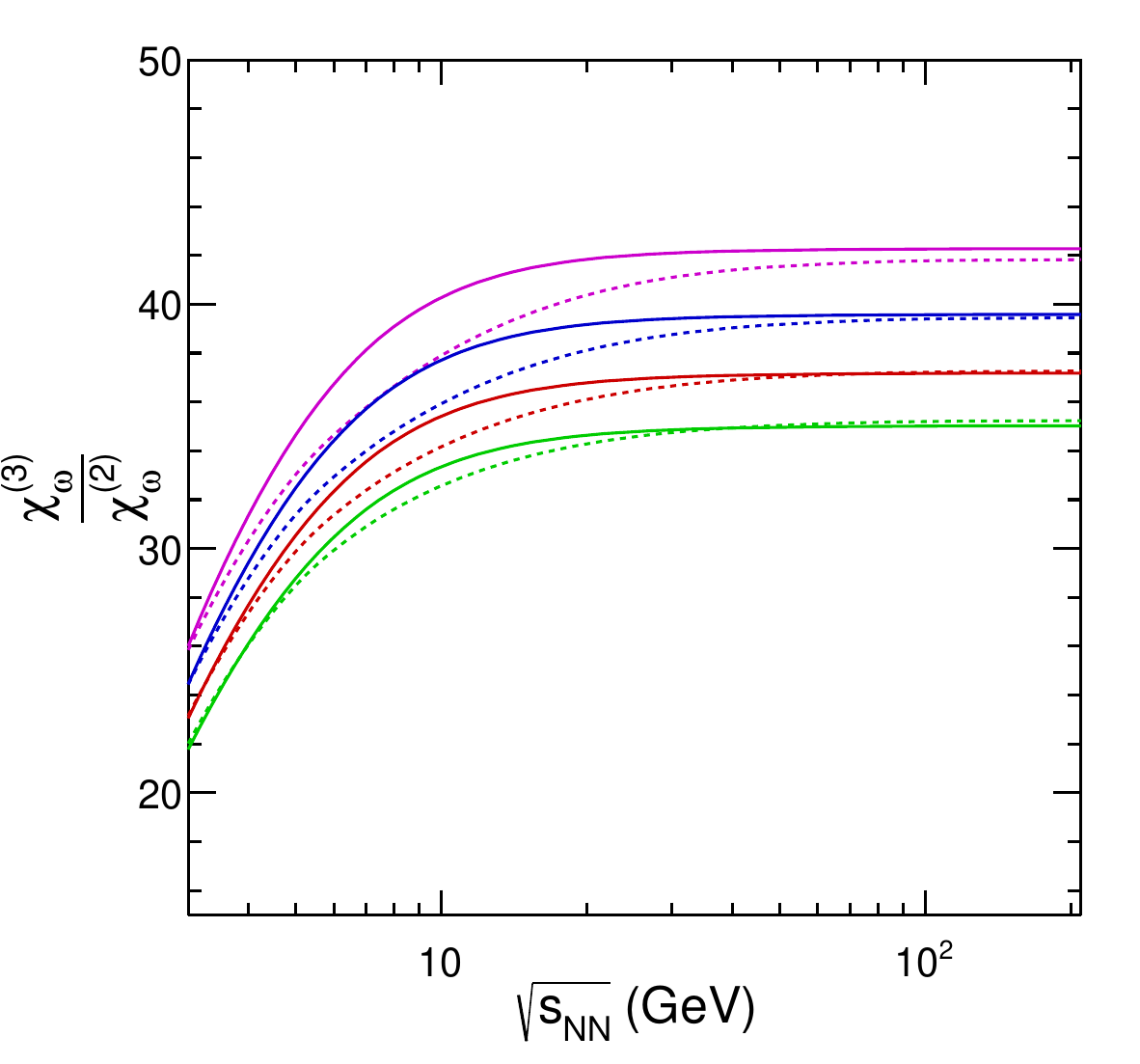}
\includegraphics[scale=0.29]{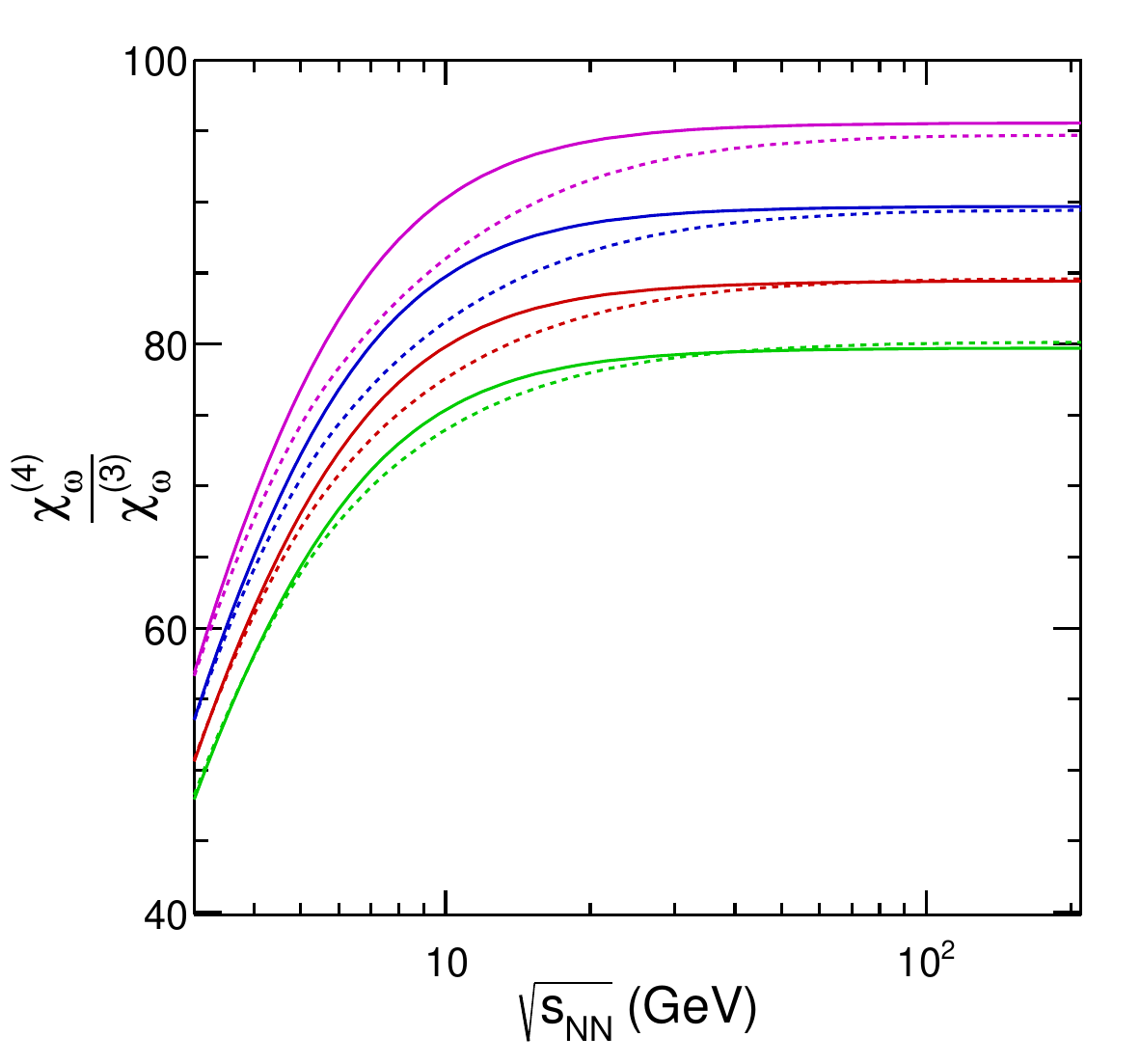}     
\includegraphics[scale=0.29]{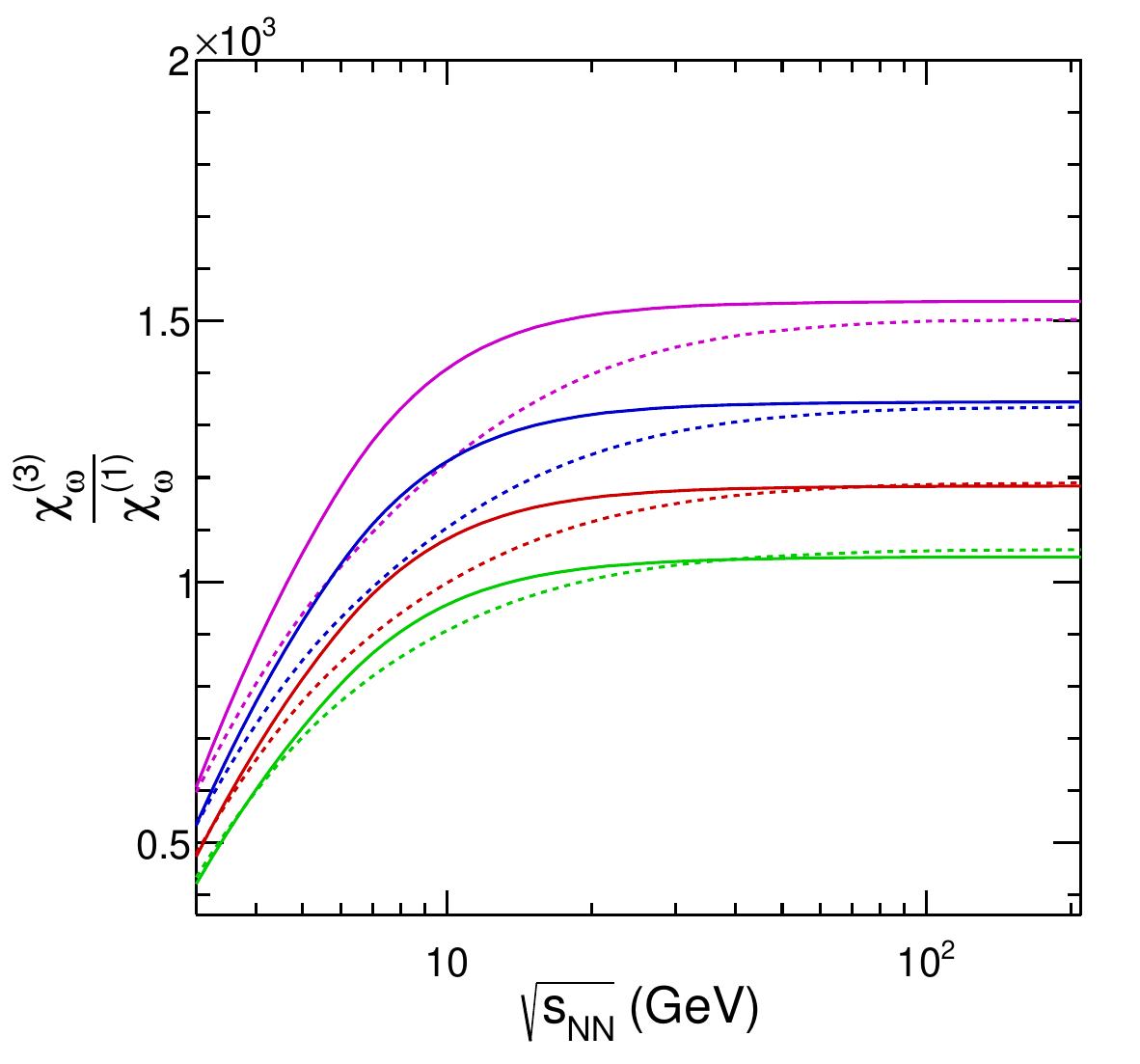}
\includegraphics[scale=0.29]{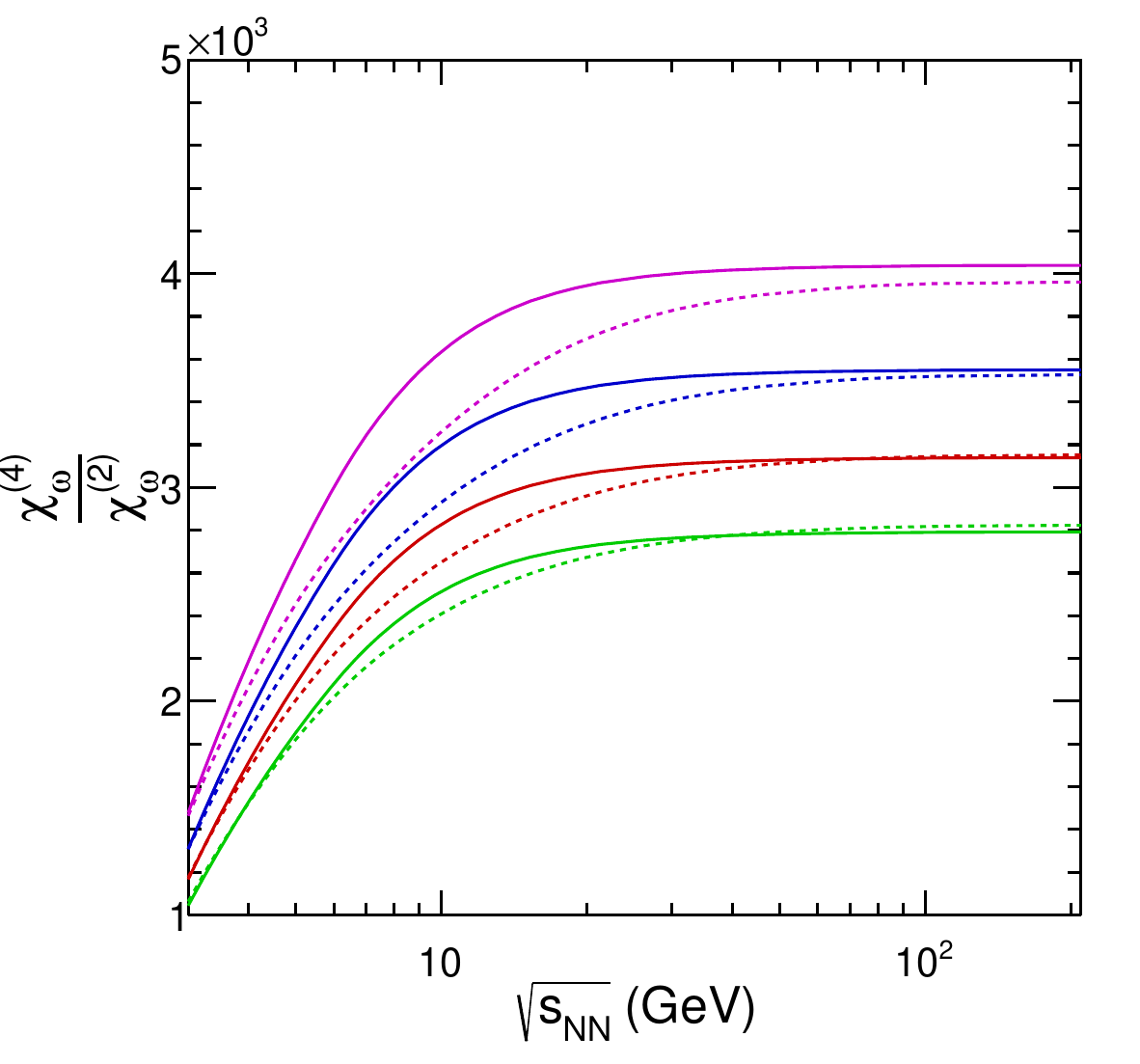}
\caption{(Color online) The variation (from left to right and downwards) of $\chi^{(2)}_\omega$/$\chi^{(1)}_\omega$, $\chi^{(3)}_\omega$/$\chi^{(2)}_\omega$, $\chi^{(4)}_\omega$/$\chi^{(3)}_\omega$, $\chi^{(3)}_\omega$/$\chi^{(1)}_\omega$, $\chi^{(4)}_\omega$/$\chi^{(2)}_\omega$ ratios as a functions of center of mass energies $\sqrt{s_{\rm NN}}$ at different values of rotation in ideal HRG (solid line) and VDWHRG (dotted line) model. A rapid change in the susceptibility ratio is observed at lower $\sqrt{s_{\rm NN}}$ ($\leq$ 20 GeV); however, a saturation behavior is observed at higher $\sqrt{s_{\rm NN}}$ ($\geq$ 20 GeV).}
\label{fig:susvssnn}
\end{figure*}

Now, we explore the rotational susceptibility ratios in the presence of finite baryon density. Figure~\ref{fig:susvsmu} shows the variation of $\chi^{(1)}_\omega$, $\chi^{(2)}_\omega$, $\chi^{(3)}_\omega$, $\chi^{(4)}_\omega$ with temperature for different values of baryochemical potential $\mu_{\rm B}$ at $\omega$ = 0.004 GeV. We set different values of $\mu_{\rm B}$ starting from 0.025 to 0.436 GeV, which correspond to the LHC, RHIC, and FAIR experiments~\cite{Braun:2001sqd, Cleymans:2006sqd, Sahoo:2023vkw}. It is expected that the strength of rotation increases with increasing $\mu_{\rm B}$ or decreasing collision energy, which can be inferred from the polarization results at STAR~\cite{STAR:2017ckg}. From Fig.~\ref{fig:susvsmu}, it is observed that all orders of rotational susceptibilities are significantly affected by baryochemical potential at lower temperatures as compared to higher temperatures. A sharp increase in the higher-order susceptibilities as a function of temperature is found compared to the first and second-order susceptibilities. At $\mu_{\rm B}$ = 0.436 GeV, all the rotational susceptibilities show a bump structure in intermediate temperature regions. Notably, the non-monotonic structure (\enquote{bump}) seen in rotational susceptibilities at low temperatures and high baryon densities mainly stems from the van der Waals equation of state, which features the nuclear liquid-gas phase transition~\cite{Vovchenko:2015vxa, Vovchenko:2015pya}. In this study, rotation serves as a probe of the equation of state rather than initiating a new or distinct phase transition. The nuclear liquid–gas phase transition is primarily density-driven and occurs without a change in the underlying degrees of freedom.
In contrast, the QCD phase transition is mainly symmetry-driven and involves a transformation of the system’s degrees of freedom from hadronic to partonic. The present framework does not establish sensitivity to deconfinement dynamics, chiral symmetry restoration property, and/or any critical QCD fluctuations near the phase transition. Any non-monotonic behavior observed here should be interpreted within the context of hadronic liquid–gas interactions encoded in the VDWHRG model, rather than as evidence of QCD critical phenomena.

Figure~\ref{fig:susratiovsmu} shows $\chi^{(2)}_\omega$/$\chi^{(1)}_\omega$, $\chi^{(3)}_\omega$/$\chi^{(2)}_\omega$, $\chi^{(4)}_\omega$/$\chi^{(3)}_\omega$, $\chi^{(3)}_\omega$/$\chi^{(1)}_\omega$, $\chi^{(4)}_\omega$/$\chi^{(2)}_\omega$ ratios as a functions of temperature $T$ for different values of baryochemical potential $\mu_{\rm B}$ at $\omega$ = 0.004 GeV. We find that these rotational susceptibility ratios follow similar trends as a function of temperature and baryon chemical potential $\mu_{\rm B}$ as the susceptibilities.  However, the peak of the bump structure at $\mu_{\rm B}$ = 0.436 GeV is relatively negligible in the ratio plots.

The thermodynamic susceptibilities (and their ratios) are related to the experimental observable cumulants/factorial cumulants (and their ratios). In this context, one can define a common useful quantity called the Binder cumulant for a rotating system, defined as follows;

\begin{equation}
U^{(4)}_{\omega} = 1 - \frac{\chi^{(4)}_{\omega}}{3 \;(\chi^{(2)}_{\omega})^2}
\end{equation}
 
In general, the Binder cumulant is widely used to characterize phase transitions and to determine the location of the critical point. The left panel of Fig.~\ref{U4} illustrates the variation of Binder cumulant $U^{(4)}_{\omega}$ as a function of temperature and rotation at zero baryochemical potential, which reflects that $U^{(4)}_{\omega}$ approaches zero at higher temperatures. While a rotational dependence of $U^{(4)}_{\omega}$ is significant at lower temperatures. Similarly, the right panel of Fig.~\ref{U4} shows the value of $U^{(4)}_{\omega}$ as a function of temperature for different baryochemical potentials at a fixed angular velocity of $\omega=0.004$ GeV within the VDWHRG model. A criticality-like behavior is observed at higher baryochemical potential around $\mu_{B} = 0.436$ GeV with finite angular velocity.

The Beam Energy Scan program at RHIC measures the net proton number (proxy for net baryon
number), net kaon number (proxy for net strangeness), and net pion, kaon, proton number (proxy for net electric charge) distribution as functions of $\sqrt{s_{\rm NN}}$ to probe the QCD critical point~\cite{STAR:2020tga, STAR:2022vlo}. Since the conserved charge fluctuations are related to the correlation length and the susceptibilities, it is worth investigating the center-of-mass energy $\sqrt{s_{\rm NN}}$  dependence of the susceptibilities that arises because of the rotation. This will provide us with an idea of how much the fluctuations due to the angular momentum are affected at a given center-of-mass energy $\sqrt{s_{\rm NN}}$. Figure~\ref{fig:susvssnn} shows the ratios of rotational susceptibilities $\chi^{(2)}_\omega$/$\chi^{(1)}_\omega$, $\chi^{(3)}_\omega$/$\chi^{(2)}_\omega$, $\chi^{(4)}_\omega$/$\chi^{(3)}_\omega$, $\chi^{(3)}_\omega$/$\chi^{(1)}_\omega$, $\chi^{(4)}_\omega$/$\chi^{(2)}_\omega$  as a functions of collision center of mass energy $\sqrt{s_{\rm NN}}$ for different values of $\omega$. To obtain the center of mass energy $\sqrt{s_{\rm NN}}$ dependence, we follow a $T-\mu_{\rm B}-\sqrt{s_{\rm NN}}$ parametrization given in Ref.~\cite{Cleymans:2006sqd}, as  
\begin{equation}
T(\mu_{\rm B}) = q_1 - q_2 \mu^{2}_{B} - q_3 \mu^{4}_{B} 
\nonumber
\end{equation}
\begin{equation}
\mu_{\rm B} (\sqrt{s_{\rm NN}}) = \frac{q_4}{1+q_5 \sqrt{s_{\rm NN}}}
\nonumber
\end{equation}

Here, $q_1$ =  0.166 GeV, $q_2$ = 0.139 GeV$^{-1}$, $q_3$ = 0.053 GeV$^{-3}$, $q_4$ = 1.308 GeV, and $q_5$ = 0.273 GeV$^{-1}$. These parameters are obtained using freeze-out criteria based on the ideal HRG model. While similar estimations have been made using the HRG model with VDW interactions and the excluded volume HRG model, the variations in the parameters are negligible~\cite{Behera:2022nfn, Tiwari:2011km, Pradhan:2023etz}.

From Fig.~\ref{fig:susvssnn}, we observe that all these ratios increase significantly at lower $\sqrt{s_{\rm NN}}$ values and start saturating towards the intermediate and high $\sqrt{s_{\rm NN}}$ ($\geq$ 20 GeV). The saturation behavior in the VDWHRG model happens towards a slightly higher center of mass energy. The effect of $\omega$ on susceptibility ratios is found to be prominent for $\sqrt{s_{\rm NN}}$ $\geq$ 20 GeV. Although obtaining the direct relation between angular momentum with the center of mass energy $\sqrt{s_{\rm NN}}$ and rotation $\omega$ is non-trivial, because it depends on several factors, such as: how much fraction of the total deposited initial orbital angular momentum is transferred to the medium (not known yet), local vortices structure, thermal gradients, medium dynamics, etc. However, it is expected that at higher center of mass energy, the fluctuations will be higher because the generated initial orbital angular momentum is directly proportional to the center of mass energies $\sqrt{s_{\rm NN}}$.

It is noteworthy to mention that, in the presence of diffusion and smearing effects, rotational susceptibilities may offer a probe of the QCD phase transition compared to traditional conserved charge susceptibilities, such as the baryon number susceptibility. Baryon number fluctuations are sensitive to local diffusion and freeze-out dynamics, making higher-order cumulants susceptible to thermal noise and experimental resolutions. On the other hand, rotational susceptibilities are tied to the system’s global angular momentum and collective vorticity. These rotational properties are less prone to being washed out by local diffusion processes, as they reflect coherent motion across the medium rather than individual particle number variations. Furthermore, rotational susceptibility, defined via the thermodynamic response of pressure to angular velocity, is less sensitive to local smearing and non-equilibrium effects because it characterizes a global thermodynamic response to rotation rather than local density fluctuations. This makes it a potentially cleaner observable. The spin polarization $\langle s^{\mu}\rangle$ obtained from experiments is proportional to spin-vorticity coupling ($\langle s^{\mu}\rangle ~\sim ~\chi_{\rm \omega}~.~\omega^{\mu}$)~\cite{Li:2020eon}, thus, by comparing polarization data to hydrodynamic models with fluid vorticity, one can, in principle, constrain the rotational susceptibility. Thus, rotational susceptibilities provide a promising and complementary avenue for probing the QCD phase structure in high-energy heavy-ion collisions. However, for rigid-body-like rotation, as in the current study, such an analogy is non-trivial because spin polarization is linked to the fluid vorticity rather than the global rotation. 


\section{Summary}
\label{sum}
For the first time, we estimate the rotational susceptibilities, quantifying the system's response to rotation, in a rotating hadron resonance gas produced in ultra-relativistic heavy-ion collisions. We evaluate these 
rotational susceptibilities and their ratios in both ideal and interacting HRG models, with the latter incorporating van der Waals interactions to simulate realistic hadronic interactions. The study shows that higher-order rotational susceptibilities are particularly sensitive to rotation.
The influence of hadron spin, system size, and baryon chemical potential on these susceptibilities is also explored as a function of temperature and rotation. In this study, rotation is used as a probe for studying the non-monotonic behavior in susceptibility ratios at low temperatures and high baryon densities, mainly stemming from the van der Waals equation of state, which features the nuclear liquid-gas phase transition. These findings highlight rotational susceptibility as a promising observable for probing QCD thermodynamics in heavy-ion collisions. In addition, such studies involving rotation exploring the phase diagram are essential and timely. 

\section*{Acknowledgement}
B.S. and K.K.P. acknowledge the financial aid from the Council of Scientific and Industrial Research (CSIR) and the University Grants Commission (UGC), Government of India, respectively. The authors gratefully acknowledge the DAE-DST, Government of India, funding under the mega-science project “Indian Participation in the ALICE experiment at CERN” bearing Project No.
SR/MF/PS-02/2021-IITI (E-37123). B.S. acknowledges Arpan Das for fruitful discussions during the preparation of the manuscript.

\end{document}